\documentstyle[twoside]{article}

\catcode`\@=11
\long\def\@makefntext#1{ 
\protect\noindent \hbox to 3.2pt {\hskip-.9pt
$^{{\eightrm\@thefnmark}}$\hfil}#1\hfill} 

\def\thefootnote{\fnsymbol{footnote}}
 \def\@makefnmark{\hbox to 0pt{$^{\@thefnmark}$\hss}}  

\def\ps@myheadings{\let\@mkboth\@gobbletwo
\def\@oddhead{\hbox{} 
\rightmark\hfil\eightrm\thepage}
\def\@oddfoot{}\def\@evenhead{\eightrm\thepage\hfil 
\leftmark\hbox{}}\def\@evenfoot{}
\def\sectionmark##1{}\def\subsectionmark##1{}}

\oddsidemargin=\evensidemargin
\renewcommand{\thefootnote}{\fnsymbol{footnote}}

\newcounter{sectionc}\newcounter{subsectionc}\newcounter{subsubsectionc}
\renewcommand{\section}[1] {\vspace{12pt}\addtocounter{sectionc}{1}
\setcounter{subsectionc}{0}\setcounter{subsubsectionc}{0}\noindent
	{\tenbf\thesectionc. #1}\par\vspace{5pt}}
\renewcommand{\subsection}[1] {\vspace{12pt}\addtocounter{subsectionc}{1}
	\setcounter{subsubsectionc}{0}\noindent
	{\bf\thesectionc.\thesubsectionc. {\kern1pt \bfit #1}}\par\vspace{5pt}}
\renewcommand{\subsubsection}[1] {\vspace{12pt}\addtocounter{subsubsectionc}{1}
	\noindent{\tenrm\thesectionc.\thesubsectionc.\thesubsubsectionc.
	{\kern1pt \tenit #1}}\par\vspace{5pt}}
\newcommand{\nonumsection}[1] {\vspace{12pt}\noindent{\tenbf #1}
	\par\vspace{5pt}}

\newcounter{appendixc}
\newcounter{subappendixc}[appendixc]
\newcounter{subsubappendixc}[subappendixc]
\renewcommand{\thesubappendixc}{\Alph{appendixc}.\arabic{subappendixc}}
\renewcommand{\thesubsubappendixc}
	{\Alph{appendixc}.\arabic{subappendixc}.\arabic{subsubappendixc}}

\renewcommand{\appendix}[1] {\vspace{12pt}
        \refstepcounter{appendixc}
        \setcounter{figure}{0}
        \setcounter{table}{0}
        \setcounter{lemma}{0}
        \setcounter{theorem}{0}
        \setcounter{corollary}{0}
        \setcounter{definition}{0}
        \setcounter{equation}{0}
        \renewcommand{\thefigure}{\Alph{appendixc}.\arabic{figure}}
        \renewcommand{\thetable}{\Alph{appendixc}.\arabic{table}}
        \renewcommand{\theappendixc}{\Alph{appendixc}}
        \renewcommand{\thelemma}{\Alph{appendixc}.\arabic{lemma}}
        \renewcommand{\thetheorem}{\Alph{appendixc}.\arabic{theorem}}
        \renewcommand{\thedefinition}{\Alph{appendixc}.\arabic{definition}}
        \renewcommand{\thecorollary}{\Alph{appendixc}.\arabic{corollary}}
        \renewcommand{\theequation}{\Alph{appendixc}.\arabic{equation}}
        \noindent{\tenbf Appendix \theappendixc #1}\par\vspace{5pt}}
\newcommand{\subappendix}[1] {\vspace{12pt}
        \refstepcounter{subappendixc}
        \noindent{\bf Appendix \thesubappendixc. {\kern1pt \bfit #1}}
	\par\vspace{5pt}}
\newcommand{\subsubappendix}[1] {\vspace{12pt}
        \refstepcounter{subsubappendixc}
        \noindent{\rm Appendix \thesubsubappendixc. {\kern1pt \tenit #1}}
	\par\vspace{5pt}}

\newcommand{\textlineskip}{\baselineskip=13pt}
\newcommand{\smalllineskip}{\baselineskip=10pt}

\def\eightcirc{
\begin{picture}(0,0)
\put(4.4,1.8){\circle{6.5}}
\end{picture}}
\def\eightcopyright{\eightcirc\kern2.7pt\hbox{\eightrm c}}

\newcommand{\publisher}[2]{{\begin{center}\eightrm\smalllineskip
	Received #1\\
	\end{center}
	}}

\def\abstracts#1#2#3{{
	\centering{\begin{minipage}{4.5in}\baselineskip=10pt\eightrm
	\centerline{ABSTRACT}
	\parindent=0pt #1\par
	\parindent=15pt #2\par
	\parindent=15pt #3
	\end{minipage} }\par}}

\newcommand{\bibit}{\nineit}

\renewenvironment{thebibliography}[1]			
	{\ninerm
	 \baselineskip=11pt				
	 \begin{list}{\arabic{enumi}.}
	{\usecounter{enumi}\setlength{\parsep}{0pt}
	 \setlength{\leftmargin 17pt}{\rightmargin 0pt}	
	 \setlength{\itemsep}{0pt} \settowidth		
	{\labelwidth}{#1.}\sloppy}}{\end{list}}

\newcounter{itemlistc}
\newcounter{romanlistc}
\newcounter{alphlistc}
\newcounter{arabiclistc}

\newcommand{\fcaption}[1]{
        \refstepcounter{figure}
        \setbox\@tempboxa = \hbox{\eightrm Fig.~\thefigure. #1}
        \ifdim \wd\@tempboxa > 5in
           {\begin{center}
        \parbox{5in}{\eightrm \smalllineskip Fig.~\thefigure. #1 }
            \end{center}}
        \else
             {\begin{center}
             {\eightrm Fig.~\thefigure. #1}
              \end{center}}
        \fi}

\newcommand{\tcaption}[1]{
        \refstepcounter{table}
        \setbox\@tempboxa = \hbox{\eightrm Table~\thetable. #1}
        \ifdim \wd\@tempboxa > 5in
           {\begin{center}
        \parbox{5in}{\eightrm\smalllineskip Table~\thetable. #1 }
            \end{center}}
        \else
             {\begin{center}
             {\eightrm Table~\thetable. #1}
              \end{center}}
        \fi}

\def\@citex[#1]#2{\if@filesw\immediate\write\@auxout	
	{\string\citation{#2}}\fi			
\def\@citea{}\@cite{\@for\@citeb:=#2\do			
	{\@citea\def\@citea{,}\@ifundefined		
	{b@\@citeb}{{\bf ?}\@warning
	{Citation `\@citeb' on page \thepage \space undefined}}
	{\csname b@\@citeb\endcsname}}}{#1}}

\newif\if@cghi
\def\cite{\@cghitrue\@ifnextchar [{\@tempswatrue
	\@citex}{\@tempswafalse\@citex[]}}
\def\citelow{\@cghifalse\@ifnextchar [{\@tempswatrue
	\@citex}{\@tempswafalse\@citex[]}}
\def\@cite#1#2{{$\null^{#1}$\if@tempswa\typeout
	{IJCGA warning: optional citation argument
	ignored: `#2'} \fi}}

\def\pmb#1{\setbox0=\hbox{#1}
	\kern-.025em\copy0\kern-\wd0
	\kern.05em\copy0\kern-\wd0
	\kern-.025em\raise.0433em\box0}

\def\fnt#1#2{\footnotetext{\kern-.3em
	{$^{\mbox{\scriptsize #1}}$}{#2}}}

\def\fpage#1{\begingroup
\voffset=.3in
\thispagestyle{empty}\begin{table}[b]\centerline{\footnotesize #1}
	\end{table}\endgroup}

\def\runninghead#1#2{\pagestyle{myheadings}
\markboth{{\eightit{\quad #1}}\hfill}{\hfill{\eightit{#2\quad}}}}

\font\tenbf=cmbx10
\font\tenit=cmti10
\font\tenit=cmti10
\font\bfit=cmbxti10 at 10pt
 1
 1
 1

\font\ninerm=cmr9
\font\nineit=cmti9

\font\eightrm=cmr8
\font\eightit=cmti8

\def\qed{\hbox{${\vcenter{\vbox{                          
   \hrule height 0.4pt\hbox{\vrule width 0.4pt height 6pt
   \kern5pt\vrule width 0.4pt}\hrule height 0.4pt}}}$}}

\runninghead{M. Janssen} {Multifractal Analysis $\ldots$}

\def\equ#1#2{
  \begin{equation} \label{#1}
    #2
  \end{equation}
}
\def\equa#1#2{
  \begin{eqnarray} \label{#1}
    #2
  \end{eqnarray}
}

\def\fig#1#2#3
{
\begin{figure}
\vspace{#1cm}
\fcaption{#2} \label{#3}
\end{figure}
}

\def\BRA{\left\langle}

\def\KET{\right\rangle}

\def\LK{\left(}

\def\RK{\right)}

\def\LBK{\left\lbrack}

\def\RBK{\right\rbrack}

\def\LB{\left\lbrace}

\def\RB{\right\rbrace}
\def\Tr{{\rm Tr}\,}

\def\C{\, \raise 2pt\hbox{$\scriptscriptstyle |$}\mskip -8mu C}

\def\ableit#1#2{\frac{\partial{#1}}{\partial{#2}}}

\begin{document}
\normalsize\textlineskip
{\thispagestyle{empty}
\setcounter{page}{1}

\renewcommand{\thefootnote}{\fnsymbol{footnote}} 


\vspace*{0.88truein}

\fpage{1}
\centerline{\bf MULTIFRACTAL ANALYSIS}
\vspace*{0.035truein}
\centerline{\bf  OF BROADLY DISTRIBUTED OBSERVABLES}
\vspace*{0.035truein}
\centerline{\bf AT CRITICALITY}
\vspace{0.37truein}
\centerline{\footnotesize MARTIN JANSSEN}
\vspace*{0.015truein}
\centerline{\footnotesize\it
 Institut f\"ur Theoretische Physik, Universit\"at zu K\"oln,
Z\"ulpicher Str. 77, 50937 K\"oln, Germany }
\vspace{0.225truein}
\publisher{}{}

\vspace*{0.21truein}
\abstracts{\noindent The multifractal analysis of disorder induced
localization-delocalization transitions is reviewed.
Scaling properties of this transition  are generic for multi parameter
coherent systems which show broadly distributed observables at
criticality.
The multifractal analysis of local measures is extended to more
general observables including scaling variables such as the
conductance in the localization problem. The relation of multifractal
dimensions to critical exponents such as the order parameter exponent
$\beta$ and the correlation length exponent $\nu$ is investigated. We
discuss a number of scaling relations between spectra of critical
exponents, showing that all of the critical exponents necessary to characterize
the critical phenomenon can be obtained within the generalized
multifractal analysis.  Furthermore we
show how bounds for the correlation length exponent
$\nu$ are obtained by the typical  order
parameter exponent $\alpha_0$ and make contact between the
multifractal analysis and the finite size scaling approach in 2-d by
relying on conformal mapping arguments.}{}{}

\vspace*{-3pt}\textlineskip
\section{Introduction}\label{secint}
\noindent
Anderson localization$^{1,2}$ has gathered interest for over three
decades. Electrons in disordered conductors can undergo a transition
to an insulating state. At zero temperatures
the diffusion constant is a function of the Fermi energy
$\varepsilon_F$
and takes a finite value in the conducting phase while it vanishes in
the insulating phase which is reached by crossing a certain value
$\varepsilon_F=E_c$ called the critical energy. This transition is due
to a disorder induced localization-delocalization transition of the
electrons wave functions. Localization occurs for strong enough
disorder
because of quantum interference effects brought about by the
randomness
of the disorder which is assumed to be static (quenched disorder).$^2$

The modeling of disorder induced localization-delocalization
transitions  (LD transitions) refers
to independent electrons of effective mass $m$ moving in a
random one-particle potential $V(\vec{r})$, i.e. the Hamiltonian
is $H=\frac{\pi^2}{2m}+V(\vec{r})$ where $\pi$ is the kinetic momentum
which may include
 magnetic fields. The potential
is characterized by
its mean value $\BRA V\KET$ and a finite range correlation function
$\BRA V(\vec{r})V(\vec{r}\ ')\KET$. A corresponding tight-binding version
with on-site random energies and nearest-neighbor hopping is referred
to as Anderson model.

Transport properties of electrons are related to two-particle Green
functions
which are able to describe the transition probability from point
$\vec{r}$ to point $\vec{r}\ '$ at a given energy.$^3$ Here the notion of
an $n$-particle Green function is introduced for the disorder average
of a product of $n$ Green functions involving retarded, $G^{+}$, and
advanced, $G^{-}$,
Green functions. These are defined as the matrix elements
of the resolvent operators corresponding to $H$, i.e.
$G^{\pm}(E):=\LK E-H\pm i\epsilon\RK^{-1}$ for $\epsilon \to +0$.

For quasi $1$-d systems (characterized by their length being much larger
then their width) rigorous results are available$^{4}$, saying that
all states become localized as the length grows infinite.
For localized states, the two-particle Green function shows
exponential decay,
$|G^{+}(E, \vec{r}, \vec{r}\ ')|\propto \exp{-|\vec{r}-\vec{r}\
'|/\xi(E)}$,
from which a {\em localization length} $\xi(E)$ can be identified.

The lower critical dimension of the LD transition is believed to be
$d=2$.$^{5,6}$ While in the absence of strong magnetic fields or
spin-orbit scattering most indications$^{5,7}$ tell that all states
are localized in the thermodynamic limit of $2$-dimensional systems
 (though the localization
length can be quite large),  $2$-d systems with strong magnetic
fields
undergo LD transitions which are believed to be responsible
for the occurrence of the quantum Hall effect$^{8,9}$ (QHE) and are
referred
to as quantum Hall systems (QHS). There is now also striking numerical
evidence
that $2$-d systems with strong spin-orbit scattering show an LD
transition.$^{10,11,12,13}$

In $3$-dimensional disordered conductors the LD transition is believed
to occur more generally$^{5}$ which has been confirmed
numerically.$^{14,15,16,17}$

Most studies on the LD transition (e.g. Refs.~$14,15,17$) were focused
on the determination of the critical value of the Fermi energy, $E_c$,
(or of the disorder strength) and on the critical exponent $\nu$ of
the localization length $\xi$. Since the late $80$'s, however, it became
clear
that the wave functions at criticality are multifractal
measures leading to a whole spectrum of critical
exponents.$^{17,18,19,20,21,22}$
 Nevertheless,  only rare connection between these new exponents
and the localization length exponent $\nu$ was made.$^{17,20,23}$

In this article, we wish to emphasize that the LD transition can be
viewed
as a prototype for a class of critical phenomena characterized by
broadly distributed observables at criticality. For this class the
multifractal analysis provides a framework to investigate all of
the critical exponents. Though, in the following, we often will refer
to the LD transition for the sake of concreteness,  we hope to make
recognition of the generalities possible (a related approach has been put
forward by  Ludwig$^{24}$ and by Ludwig and Duplantier$^{25}$).

The multifractal analysis is a scaling approach relying on the
principle
of ``absence of length scales''.
Systems without characteristic intrinsic length scales
(defining the interrelations
of the systems constituents)  obey
homogeneity  laws with respect to rescaling.
Let $A$ be an operator   or a complex valued function defined for
values $x$  describing a length scale of descriptive nature (e.g.
 wavelength, system size). Absence of length scales means
that $A$  shows a typical homogeneity
law
\equ{11.1}
{
	A(s\cdot x) = s^{\kappa}\cdot A(x)
}
where $\kappa$ is called the homogeneity exponent and $s$ is a real
number. In other words: the absence of length scales is reflected by
the property that a rescaling of $x$ can be compensated by a
rescaling of the observable $A$. For real valued functions $A(x)$ the
solution of the homogeneity equation (\ref{11.1}) is a power law:
$A(x)\propto x^{\kappa}$.
Now, assume that $A^{\lbrack q\rbrack}(x)$ is a functional of powers
$q$
of those observables which are involved in the definition of $A(x)$,
then the simplest situation that may appear is that $\kappa(q)$,
defined by
\equ{11.2}
{
	A^{\lbrack q\rbrack}(s\cdot x)= s^{\kappa(q)}\cdot A^{\lbrack
	q\rbrack}(x)\, ,
}
is a linear function of $q$; the operators simply add
their exponents. If  $\kappa(q)$ shows a
significant deviation from linearity we say that the scaling behavior
of $A(x)$ is {\em anomalous}.  Multifractality will turn out to be a
generic case of anomalous scaling behavior.

The article is organized as follows. In Sec.~$2$
several aspects of critical phenomena, especially the notion of order
parameters, correlation lengths and critical exponents, are reviewed
and the LD transition is put into this perspective. It turns out that
the LD transition shows unusual features, the most important of which
is the occurrence of a dependence of the local susceptibility on the
system size.
In Sec.~$3$ we review the multifractal analysis of measures
and the notion of $f(\alpha)$ spectra describing the scaling of broad
distributions for local probabilities. Results for the wave function
(defining a local measure)
of QHS are discussed in detail.
Generalizing to positive box-observables in Sec.~$4$ will allow
us to relate the critical exponent of the correlation length, $\nu$,
to the multifractal exponents $\alpha_0$ (the {\em typical scaling exponent}
of box-probabilities)  and $X$, the  {\em normalization exponent}
of
a scaling variable. Furthermore, we discuss
 bounds for $\nu$ in terms of multifractal exponents.
In Sec.~$5$ we focus on correlations in local box-observables.
This topic is touched already in previous sections, but is then
treated
more systematically. As a result we find scaling relations which
determine all of the correlation exponents by the original
$f(\alpha)$-spectrum
and the corresponding normalization exponents. By relying on conformal
mapping arguments we conjecture scaling relations to determine the
multifractal spectra via finite size scaling methods in $2$-d.
Sec.~$6$ contains our conclusions.

\section{Aspects of Critical Phenomena}\label{secasp}
\subsection{Critical Exponents}\label{subcri}
\noindent
A  defining feature
of a critical phenomenon is the divergence of  the {\em{correlation
length}} $\xi_c$, defined by the spatial decay  of the statistical
correlations
of a local field,
\hbox{$
\BRA \varphi(0)\varphi(r)\KET \propto \exp\LK-r/\xi_c\RK $},
has
to diverge at the critical point  $T_c$.
In the sequel of this article
we refer to the following nomenclature for
critical phenomena.
The physical states are described by state coordinates (we will refer
to them by one symbol $T$) and by order parameters ($m$). For infinite
system sizes the state coordinates exhibit a critical point $T_c$
where the order parameters vanish with a power law in $t:=|T-T_c|$
\equ{8.7}
{ 	m  =  0 \; (T < T_c)\;\; ,\;\;\;
m  \propto
t^{\beta}   \; (T > T_c)\, .
 }
There exists a generating function $F(t, h)$ where $h$ is the
conjugate field with respect to the order parameter
\hbox{$ 	m=\ableit{F}{h} (h\to 0) $ }.
In analogy to equilibrium
systems we call $F$ the free energy per unit volume. Higher
derivatives of $F$ also show power law behavior for $h=0$, $t\to
0$. For example
\equ{8.77}
{ 	\chi = \frac{\partial^2 F}{\partial h^2} (h\to 0) \propto
t^{-\gamma} \; ,\;\; c = \frac{\partial^2 F}{\partial t^2} (h\to 0) \propto
t^{-\alpha}\, .
}The exponents $\gamma$ and $\alpha$ (the latter  may be negative)
are called critical exponents of the susceptibility, $\chi$, and specific heat,
 $c$, respectively.

To have a concept of spatial correlations we assume that the order
parameter and its conjugate are local fields $m(\vec{r})$,
$h(\vec{r})$ in a $d$-dimensinal system of volume $V$
for which a generating functional
\equ{8.11}
{ 	Z\LK h(\vec{r}), t\RK = \int \, d  [\varphi(\vec{r})] \exp
\LB - {\cal{H}}\LK [\varphi(\vec{r})], h(\vec{r}); t \RK\RB
} exists where ${\cal{H}}$ is commonly called Hamiltonian
although it does not have to be hermitean.
The functional $F(h,t)$
is given as $F(h,t)=-\ln Z(h,t)/V$ for the volume $V\to
\infty$.  While the mean value $m$ of the order parameter is given by
the first moment of the generating functional,
$
 	m  =  \BRA \varphi(\vec{r}) \KET
$, correlation functions are given by higher
order cumulants
\equ{8.14}
{ 	\BRA \varphi(\vec{r_1})\ldots \varphi(\vec{r_n})\KET_C :=
\frac{\delta^n \ln Z}{\delta 	h(\vec{r_1})\ldots h(\vec{r_n})}(h\to
0)\, .  }
 which are related to  moments in a  combinatoric
way, for example:
{$ \BRA \varphi(\vec{r})\KET_C  =  \BRA \varphi(\vec{r})\KET$}
and
\equ{8.1444}
{ 	\BRA \varphi(\vec{r})\varphi(\vec{r}\ ')\KET_C  =  \BRA
\varphi(\vec{r})\varphi(\vec{r}\ ')\KET - \BRA \varphi(\vec{r})\KET \BRA
\varphi(\vec{r}\ ')\KET=: \chi(|\vec{r}-\vec{r}\ '|) \, . }
$\chi(r)$ is called local susceptibility, since the generating
formalism allows to conclude \hbox{$\int d^d r\, \chi(r) = \chi$}.
 The correlation length $\xi_c$ can  be identified from
\hbox{$\chi(r)\propto \exp{-r/\xi_c}$} and is assumed to diverge with
a power law approaching $t=0$,
\equ{8.15b}
{
	\xi_c(t)\propto t^{-\nu} \, ,
}
where $\nu$ is called critical exponent of the correlation length.

To illustrate the analogies of the LD transition with critical
phenomena as outlined here we refer to the fact$^3$ that the LD
transition is characterized by the analytic behavior of the disorder
averaged two-particle Green function,
\equ{green}
{
g^2(|\vec{r}-\vec{r}\ '|)=\overline{<\vec{r}|G^{+}|\vec{r}\ '>
<\vec{r}\ '|G^{-}|\vec{r}>}\, ,\nonumber
}
with respect to the infinitesimal parameter $\epsilon$ in
$G^{\pm}=(E-H\pm i\epsilon)^{-1}$ which distinguishes retarded and
advanced Green functions and controlls the long time averages
\index{long time averages} of transport quantities such as the
diffusion constant $D(E)$,
\equ{diffuconst}
{ 	\rho(E) D(E) = \lim_{\epsilon \to 0^{+}}
 \frac{\epsilon}{2\pi\hbar} 	\epsilon   \int d^d
r\, r^2 g^2 (r)\,    ,
}
where \hbox{$\rho(E)=\frac{1}{2\pi
i}\overline{<\vec{r}|G^{-}(E)-G^{+}(E)|\vec{r}>}$} is the average
density of states.
Thus, it seems natural that a critical phenomenon description of the
LD transition may start from a generating function for disorder
averaged
Green functions.
 For our purposes we do not need to go
into technical details and leave the explicit form of the
corresponding Hamiltonian open.
We only like to refer to some of the
main features of such theories (for details see e.g. Refs.~$26,27$)
 which can be
 summarized as follows.

The degrees of freedom are matrix fields $Q$ with a block structure
{$Q^{pp'}$ where $p,p' \in \LB\pm\RB$ correspond to
retarded ($+$) and advanced ($-$) Green functions, respectively.
The source field is the infinitesimal parameter $\epsilon$
and the corresponding source term in the Hamiltonian is
$
	\epsilon\LK Q^{++}-Q^{--}\RK
$
leading to
\equ{critdensity}
{
	\frac{\delta \ln Z}{\delta\epsilon}=\BRA Q^{--}\KET - \BRA
Q^{++}\KET = \overline{<r|G^{-}|r>-<r|G^{+}|r>}
}
which is proportional to the average density of states $\rho(E)$.
Thus, the density of states in disordered electron systems seems to be
a candidate for an order parameter of the LD transition.
Unfortunately,
the density of states is a smooth function of energy$^{3,28}$
unable to reflect the LD transition which is characterized by the
vanishing
of the diffusion constant related to the  susceptibility via
Eqs.~(\ref{green}, \ref{diffuconst}).
Before coming  back to this peculiarity of the LD transition let
 us
 proceed in listing common knowledge about scaling ideas
applied to critical phenomena.

\subsection{Scaling Relations}\label{subscal}
\noindent
The crucial assumption of any scaling approach to critical phenomena
(see e.g. Ref.~${29}$)
is that the critical
exponents $ \alpha$, $\beta$, $\gamma$ and $\nu$ have their origin in
the divergence of one relevant length scale, the correlation length
$\xi_c$.
Since thermodynamic quantities are generated by the free energy we
make the homogeneity assumption that the free energy, in the vicinity of
the critical point, can be written as
\equ{8.3.1}
{
	F(t ; h) = a t^{\Box} f_{\pm}( h/t^{\Delta}) +f_r
}
where $f_{\pm}$ is a regular function of the argument
 $h/t^{\Delta}$ and $f_r$ stands for regular terms which do not show
scaling behavior for  $h\to 0$, $t\to 0$.
By differentiating $F$,  the following scaling
relations\index{scaling relation}
\equ{8.3.2}
{
	2\beta +\gamma = 2-\alpha\;\; ,\;\;\; \Box =2-\alpha\,
}
can be concluded.
The scaling exponent $\nu$ of the correlation length $\xi_c$ which
relates the divergence of $\xi_c$ with the vanishing of $t$ is
related to these exponents by the following scaling argument.
 Consider a finite system $\Omega$
of volume $L^d$ for which the
free energy per unit volume is
 \equ{freeperu}
{
	F_L(t,h)=-\frac{\ln Z_L(t, h)}{L^d}\, .
}
A change of $L$ in $Z_L(t_0,h_0)$ to $L'=\lambda L$ for fixed values $t_0, h_0$
is expected to  be compensated by an appropriate change in $t, h$
 such that the
partition sum remains unchanged, i.e.
\equ{partinv}
{
	Z_L(t_0,h_0)=Z_{L'}(t',h')\, .
}
This procedure defines implicitly functions $t'(L,L'),h'(L,L')$
with $t(L,L)=t_0$, $h(L,L)=h_0$.
Consequently,
\equ{freescale}
{
	F_L(t_0,h_0)=F_{L'}(t',h')\cdot\lambda^d\, .
}
Choosing the value $L=\xi_c(t_0)$
we see
\hbox{$F_{\xi_c(t_0)}(t_0,h_0)=F_{L'}(t',h')\cdot\xi_c^{-d}
(t_0)$}
and find a scaling relation\index{scaling relation}
\equ{nud}
{
	\Box=d\cdot\nu\, .
}
We can  cast the scaling relations Eq.~(\ref{8.3.2})
in a form, where only $\alpha$, $\beta$, $\gamma$ and $\nu$ appear
\equ{scalrelation}
{
	2\beta +\gamma =2 -\alpha = d\cdot\nu\, .
}
This means that only two of these four exponents are independent.

We introduced the correlation length with the help of the local
susceptibility. At the critical point, where $\xi_c$ diverges, the
correlation
 function can show a homogeneity law
\footnote{Note,  that in the literature on critical phenomena the exponent
$\tilde{z}$ is often written in a form ${\tilde{z}}=d-2+\eta$. We will
avoid this notation since it could lead to confusions in the context
of the LD transition where an exponent $\eta$ is used sometimes in a different
context (see the discussion below).}
\equ{8.30.1}
{
	\chi(r)\propto r^{-\tilde{z}}\, .
}
Owing to the general relation between the global susceptibility
$\chi$ and the correlation function $\chi(r)$,
\hbox{$\chi =\int d^{d}r\,\chi(r)$}, and by introducing the correlation
length
 $\xi_c(t)$ as a cut-off length in the  integral over $\chi(r)$,
which diverges at $t=0$,  we find another scaling
relation\index{scaling relation} which is
independent of the homogeneity assumptions for the free energy
\equ{8.30.2}
{
	\gamma =\nu\cdot\LK d-{\tilde{z}}\RK\, .
}
Together with the scaling relations of Eq.~(\ref{scalrelation}) this
tells
that
only two of the five exponents $\alpha$, $\beta$, $\gamma$, $\nu$ and
$\tilde{z}$ are independent.

We now come to a crucial point for what follows on the multifractal
analysis
of broadly distributed observables at criticality.
Imagine that in addition to the distance $r$
the finite system size $L$ is in a regime of  absence of
length scales.
Assume that  the
function
\hbox{$C(r,L):= \BRA \varphi(\vec{r})\varphi(\vec{r}\ ')\KET_L$}
shows a homogeneity law with respect to both lengths, $r$ and $L$,
\equ{ansatz}
{
	C(r,L)\propto r^{-z}L^{-y}\, .
}
The   exponent $y$ describes the system length dependence and
$z\not=\tilde{z}$ describes the distance dependence. That such a
situation will indeed appear in the LD transition problem will be discussed in
Sec.~$2.4$ and in more detail in Secs.~$3.3,5.2$.
A similar reasoning now
shows that
\equ{8.30.5}
{
	\gamma = \nu\cdot \LK d -(y+z)\RK\, , \; y+z ={\tilde{z}}
}
where the last relation follows from consistency with
Eq.~(\ref{8.30.2}).
Notice that identifying the exponent $\eta$ in $z=d-2+\eta$ as the
correlation exponent means that {\it such} $\eta$ is not determined by
$\gamma $ and $\nu$ alone, but requires the additional
knowledge of $y$.

\subsection{$\beta$-Functions}\label{subbet}
\noindent
A method to test scaling assumptions and
calculate critical exponents is the so-called
 renormalization group.
 The definition of the renormalization group
is not unique, but can in general be described as a transformation
acting
on the
Hamiltonian ${\cal H}$  of the system where a length scale serves as
transformation parameter. Instead of introducing a finite
system size, one can also   work in the thermodynamic limit considering the
system to be defined on a lattice. Changing
   the lattice
constant
$a_0$ to  $L_b=ba_0$,
and changing the coordinates $t ,h$ in an appropriate way,
one tries to keep  $\ln Z$ fixed.
After many iterations of this procedure (which defines the
renormalization group)
one ends up with a new Hamiltonian the structure of which should not
have changed
except for the change
in the coordinates and that the fields are now defined on a new
lattice.  That such a procedure indeed works makes the model
renormalizable. Otherwise one had to introduce  a number (in the
worst case an infinite number) of new terms and coordinates into the
new
Hamiltonian in order to keep  $\ln Z$ fixed.
Thus,
 the application of the renormalization group to the free energy
density
 leads to
\equ{8.5.1}
{
	F_0:=F(t ,h) = b^{-d} F(t' ,h')
}
where $t'=b^{y_{t}}t$, $h'=b^{y_h}h$ in the critical regime and $y_h$,
$y_t$ are  scaling exponents.
Choosing the factor $b$ such that $t'=const.$,  we
arrive again at the scaling relations\index{scaling relations}
 of Eqs.~(\ref{8.3.2},\ref{scalrelation}).
By considering the system size $L$ being the scaling parameter,
an extension of  renormalization group ideas can be made,
based on   the
following assumption.
Far away from the critical regime (where  power
law scaling holds) there still exist {\em{scaling variables}}
$\Lambda (L)$ which  fulfill a restrictive functional equation:
\equ{8.5.9}
{
	\Lambda(bL) = f\LK \Lambda (L), b\RK\, .
}
For simplicity  we consider the case of only one scaling variable here.
 This means that the
value of $\Lambda$ for system size $bL$ only depends on the value of
$\Lambda$ for system size $L$ and the scaling factor $b$ where $f$ is
called a scaling function.
This assumption is very strong since, in
general, $\Lambda$ will depend on many microscopic parameters. The idea
behind this assumption is that most of these parameters will become
unimportant for describing the flow of the scaling variable with
increasing system size, and that the scaling function $f$ is the only
information needed to determine this flow. The existence of a scaling
function $f$ is equivalent to the existence of a
$\beta$-function\index{$\beta$-function}
\equ{8.5.10}
{
	\beta\LK\ln\Lambda\RK:= \frac{d  \ln\Lambda (L)}{d  \ln
	L}=\frac{1}{\Lambda}\frac{\partial f}{\partial b}\LK\frac{\partial
	f}{\partial \Lambda}\RK^{-1} (b=1)
}
which is indeed a function of $\ln\Lambda$ alone. Provided the
$\beta$-function
is smooth, the flow is given by the solution of the differential
equation
\equ{8.5.11}
{
	\beta\LK\ln\Lambda\RK = \frac{d  \ln\Lambda}{d  \ln L}\, .
}
Eq.~(\ref{8.5.11}) is  called a renormalization group
equation. The
regime where $\beta$ can be linearized around a fixed point $\Lambda^*$ of the
flow ($\beta \LK\ln\Lambda^*\RK =0$) is called {\em critical regime},
\equ{8.5.12}
{
	\beta\LK \ln \Lambda\RK = \beta'\cdot\LK \ln \Lambda
-\ln\Lambda^*\RK\, .
}
Here $\beta'$ is the slope of the $\beta$-function at $\Lambda^{\ast}$.
Starting from a system of size $L=L_0$, with $\Lambda=\Lambda_0$ chosen
close to $\Lambda^*$, and turning on the renormalization flow until the
system reaches a size $\xi_c$ marking the bound of validity of
Eq.~(\ref{8.5.12}) (which is here by definition the {\em correlation
length}), one finds
\equ{8.5.13}
{
	\LK\frac{\xi_c}{L_0}\RK^{\beta'} =
	\left(\frac{\ln\Lambda-\ln\Lambda^\ast}{\ln\Lambda_0-\ln\Lambda^\ast}
	\right)	 		\, .
}
 The critical regime is narrow, we expand
$\Lambda(\xi_c)$ and $\Lambda_0$ around $\Lambda^*$ and get for the
correlation length
\equ{8.5.14}
{
	\xi_c =
	L_0
	\left(\frac{\Lambda-\Lambda^\ast}{\Lambda_0-
	\Lambda^\ast}\right)^{1/\beta'}  \, .
}
 Consider the situation where
the width of the critical regime
\hbox{$\Delta\Lambda=
\Lambda_0
-\Lambda^{*}$}
is triggered by the parameter $t$ (leaving $L_0$ and
$\Lambda-\Lambda^{*}$
unchanged),  the divergence
\equ{8.5.16}
{
	\xi_c \propto \LK\Delta\Lambda\RK^{-\nu}\propto  t^{-\nu}
}
follows,
where the exponent $\nu$ is given by the inverse slope of the
$\beta$-function at the fixed point, i.e.
\equ{nubeta}
{
	\nu=1/\beta'(\Lambda^{*})\, .
}

\subsection{Unusual Features of the LD
Transition}\label{subunu}
\noindent
 It is now time to put the LD transition into the perspective of the
critical phenomena terminology.
A common feature of field theoretic
approaches to the LD transition is that disorder averaged n-particle
Green functions are generated by a
field theoretic generating functional in the sense of Sec.~$2.1$.
As already mentioned,  the density of states which is a
one-particle Green function appears as the formal order parameter
and the conjugate field to the density of states is the
infinitesimal
parameter of the Green function. The density of states does not show
the LD transition, but is a smooth function of the
energy which, in the problem of
the LD transition, is the analog of the temperature in equilibrium phase
transitions.
Consequently,
the ``critical exponent'' $\beta$ of this formal order parameter is zero
\equa{8.8.1}
{
	\beta =0\, .
}
The LD transition cannot be described by the formal order
parameter, Eq.~(\ref{critdensity}), of the generating functional.
The LD transition can only be obtained by analyzing quantities which
are related to the two-particle Green function such as the diffusion
constant $D(E)$.

Furthermore,
the scaling relations, Eqs.~(\ref{scalrelation},\ref{8.30.2}),
tell  that the correlation
exponent
of the local susceptibility (a two-particle Green function) vanishes, i.e.
\equ{8.8.2}
{
	{\tilde z} =0\, .
}
This result seems to rule out a power law behavior of the
two-particle Green function (e.g. the  density correlator).
However, Chalker and Daniell$^{30}$ demonstrated,
by numerical calculations, the  density correlator to show  a power law
behavior at the LD transition of a quantum Hall system,
$\propto r^{-0.38}$. In addition, by relying on an exact inequality obtained by
Chalker$^{31}$ one has analytical evidence
 that the density
correlator at the critical point cannot be described by a vanishing distance
exponent. This phenomenon is often referred to as {\em anomalous diffusion}.
 Fortunately, from our discussion of correlation exponents in
Sec.~$2.2$ where we distinguished between different types of
distance exponents $z$ and $\tilde{z}$ we can fairly conclude
 that the value of $0.38$
does not correspond to $\tilde{z}$,  but corresponds
to the distance exponent $z$ in
a regime where the correlation function shows power law also with
respect to  another length scale which can be interpreted as a system
size $L$ being much smaller than the correlation length $\xi_c$, i.e.
\equ{ztildez}
{
	z=0.38 \not= \tilde{z}=0\, .
}
The phenomenon of anomalous diffusion and the unability of the density
of states to describe the LD transition are two of the main unusual
features of the LD transition. By Eq.~(\ref{ztildez}) it is indicated
that both are not independent of each other.

Another unusual feature of the LD transition which, at first sight,
seems to be independent of the previous ones was observed by
Altshuler et al.\,.$^{32,33}$ They
reexamined
the
phenomenological scaling theory for the LD transition of Abrahams et
al.\,.$^{33}$ In this scaling theory the
conductance $g(L)$
   of a cube with linear dimension $L$ is considered to be a scaling
 variable  in the  sense of Sec.~$2.3$,
i.e. $g$ obeys a differential equation in terms of
a $\beta$-function
\equ{8.7.1}
{
	\frac{{d }\ln g}{{d }\ln L}
	=\beta \left( \ln g\right)
}
which is a unique function of $\ln g$.
As shown by Altshuler et al.$^{32,33}$ the conductance of mesoscopic
systems highly depends  on the individual properties of a given system
(e.g.~on the concrete disorder potential). As a result an ensemble of
different systems reflects
 a broad distribution in $g$
which cannot be characterized by the mean value
of $g$ alone. Consequently, the mean value of the conductance $\BRA g\KET$
is  {\em not}  a suitable scaling variable.
Instead,
in any  scaling approach
to the LD transition
one has to consider
the whole distribution function of the conductance. However,
it may be possible to apply the ideas of the renormalization group to
certain
parameters of this distribution function.
Following Shapiro$^{34}$
we refer to an $s$-parameter scaling theory
for the LD transition  if $s$ parameters ${g_{\rm rel }}_{1,\ldots ,s}$
obeying scaling equations
\equ{8.7.2}
{
	\frac{{d }\ln {g_{\rm rel}}_\mu}
	{{d } \ln L} = \beta_\mu \left( \lbrace
	{g_{\rm rel}}_k\rbrace_{k=1,\ldots ,s}\right)
}
are required to determine the distribution function of a scaling
variable $g$.
If, at least in the vicinity of the transition point
 ($\beta_\mu \left( \lbrace
	{g_{\rm rel}^{*}}_k\rbrace_{k=1,\ldots ,s}=0\right)$),
only {\it one} relevant length scale $\xi_c$ and {\it one}
type of scaling variable ${g_{\rm rel}}(L)$
exists, then the critical exponent $\nu$
of {\it the} correlation length $\xi_c$ is defined by
\equ{8.7.3}
{
	{g_{\rm rel}}(L)-g_{\rm rel}^{*} \propto L^{1/\nu}
	\;\;\; {\rm for}\;\; L\leq \xi_c
}
and $1/\nu$ is given by the slope of the corresponding $\beta$-function
at $g_{\rm rel}^{*}$.
A context where this concept will serve to be fruitful
is provided by the generalized multifractal analysis
to be discussed in
Sec.~$4$. There we will also see that anomalous diffusion
and broad distributions of physical quantities are deeply connected.

\section{Multifractal Analysis of Measures}\label{secmul}
\noindent
After having discussed some aspects of critical phenomena, where
the absence of length scales is reflected by the existence of critical
exponents,
the multifractal analysis of measures is reviewed in this
section.$^{35,36,37,38}$  This analysis
is appropriate to describe  self similar local observables
which can be interpreted as measures. It
 can also be viewed as an
extension
 of the fractal dimension approach to self similar geometric objects
invented by Mandelbrot.$^{39}$

\subsection{Scaling of Moments}\label{submom}
\noindent
To introduce the notion of the multifractal analysis and
for the sake of concreteness
we consider a
quantum Hall system at criticality.
For a finite two dimensional quantum Hall system of linear size $L$
with double periodic boundary conditions we study
 the electrons wave function
$\psi(\vec{r})$, the modulus of which defines a normalized measure.
The probability for an electron to be found in a box of linear size $L_b$
is given by the {\em box-probability}
\equ{11.3}
{
	P(L_b):= \int\limits_{\rm box}{d }^2 r\, |\psi(\vec{r})|^2\, .
}
Covering the system by a mesh of $N(L_b,L)$ boxes the fractal dimension $D$
of the wave functions support is defined by
$N(\lambda)\propto\lambda^{-D}$
where $N(\lambda)$ is the number of boxes with non-vanishing
box-probability for a given ratio $\lambda=L_b/L$.
Since the wave function is never exactly zero in any box, $N(\lambda)$
equals
the total number of boxes $N(L_b,L)$ and, consequently, the fractal
dimension is the Euclidean dimension of the system, i.e. $D=d=2$.
At the level of the fractal dimension $D$ the wave function shows no
interesting fractal behavior. Let us now focus our attention to the
scaling behavior of the box-probability. The normalization condition
$\sum_i^{N(\lambda)} P_i(L_b)\equiv 1$
yields the scaling behavior  for the average
\equ{11.6}
{
	\BRA P(L_b)\KET_L \propto \lambda^{D}\, .
}
Here the average of  quantities $A_i$ (corresponding to the box numbers
$i$) is defined by
$\BRA A \KET_L := \frac{1}{N(\lambda)}\sum_i^{N(\lambda)} A_i\,.
$
Let us
  consider  higher moments of the
box-probability $P(L_b)$
with respect to this averaging procedure.
The general assumption underlying the multifractal analysis is
that
for a finite interval of values of $\lambda$ the moments $\BRA
P^q(L_b)\KET_L$
show power law behavior indicating the absence of length scales in
the system, i.e.
\equ{11.9}
{
	\BRA P^q(L_b)\KET_L \propto \lambda^{D+\tau(q)}
}
where $\tau(q)$ does not depend on $\lambda$. In
quantum Hall systems the assumption about the absence of length scales means
  that on the one hand
the correlation length\index{correlation
length} $\xi_{c}$ of the
LD transition
 has to be much larger than the lengths $L_b,L$. On the other hand
microscopic length scales $l$,
such as the cyclotron radius $r_c$,  have to be much smaller than $L_b,L$.
In summary the condition for the multifractal analysis to be useful
is
\equ{11.10}
{
	l \ll L_b < L \ll \xi_c.
}
For finite systems the states which are candidates for the
multifractal analysis are states with localization lengths $\xi$
being much larger than the system size. This corresponds to the
critical states of the LD transition in finite systems. In the thermodynamic
limit $L\to\infty$ such states can only be found at the critical energy
$E_c$. In this limit $\lambda$ goes to zero and the function $\tau(q)$
can be defined uniquely for the critical state by
\equ{11.11}
{
	\tau(q):= \lim_{\lambda \to 0} \frac{\ln \LK\BRA
	P^q(L_b)\KET_L\RK}{\ln \lambda} -D\, .
}
For finite systems one can give an estimate for $\tau(q)$
by considering the slope in a plot of \hbox{$\ln \LK\BRA
	P^q(L_b)\KET_L\RK$} versus $\ln \lambda$ over an interval of
values $\lambda$ which fulfill the above mentioned constraints.
Now we want to discuss how multifractality is
reflected by the properties of the function $\tau(q)$. Note that by
the construction of
\hbox{$\BRA P^q(L_b)\KET_L$} we have found an observable in the sense
of  Eq.~(\ref{11.2}) . The function $\tau(q)$ corresponds to the homogeneity
exponents $\kappa(q)$ and multifractality of the wave function means
that $\tau(q)$ is a non-linear function of $q$.  According to the
normalization condition,
$\tau(1)=0 \, , \; \tau(0)=-D$
and $\tau(q)$ can thus be parametrized by the
{\em generalized  dimensions} $D(q)$, defined by
\equ{11.13}
{
	\tau(q)=:D(q)\cdot (q-1)\, ,
}
and the fractal dimension $D=D(0)$. Consequently, a single-fractal
  wave function is characterized
completely by the fractal dimension $D$, i.e. $D(q)\equiv D$.
 $D(q)$ being  not a constant function distinguishes
multifractals from single-fractals.

The reason why  the multifractal analysis is  a powerful
method to analyse scaling behavior comes from the fact that one can
make very general statements about the analytic behavior of the
$\tau(q)$ function and the $D(q)$ function, respectively.
Assuming smoothness of both functions for $q$ being real numbers
one can derive the following results:$^{38,40}$
\begin{itemize}
\item The function $\tau(q)$ is a monotonically increasing
   function with negative
curvature, i.e.
\equ{11.14}
{
	\frac{d  \tau(q)}{d  q} > 0\, , \; \frac{{d }^2 \tau(q)}{d
	q^2}\leq 0       \, .
}
\item The generalized dimensions $D(q)$ are positive, monotonically
decreasing
 and bounded by
$D_{\pm\infty}:=D(q\to\pm\infty)$, i.e.
\equ{11.15}
{
	\frac{d  D(q)}{d  q} \leq 0\,\, , \;\; 0\leq D_{\infty} \leq D(q)
	\leq D_{-\infty}\, .
}
\item The function $\tau(q)$ has asymptotically constant slopes given by
$D_{\pm \infty}$
\end{itemize}
Here a sketch of the
 proves of these statements is given (for details see Refs.~$38,40$)
   The main part of the derivations relies on
the homogeneity, reflected  by power laws, already for finite ratio
$\lambda$,
and on the normalization condition which keeps the box
probabilities less or equal to unity.
First, the monotonicity of $\tau(q)$ is a simple consequence of
 $P_i\leq 1$ and, thus,
\hbox{$\sum_i P_i^q < \sum_i P_i^{q'}$} for $q>q'$.
A much stronger statement is the monotonicity of $D(q)$ which
can be shown by writing $D(q)$ as
\equ{11.16}
{
	D(q)=\frac{1}{1-q}\frac{\ln \LK \sum_i P_i^q\RK}{\ln \lambda}
	= \ln\LB \LBK \sum_i
	P_iP_i^{q-1}\RBK^{1/(q-1)}\RB\cdot(\ln\lambda)^{-1}
} with the help of a generalized H\"older inequality,
\equ{11.17}
{
	\LBK \sum_i P_i(P_i)^r\RBK^{1/r} \geq
	\LBK \sum_i P_i(P_i)^{r'}\RBK^{1/r'} \, \;\; {\rm for}\;\;\;
	r>r'\, .
}
  Notice that for
distinct values of $q$ and $q'$ the equality
$D(q)=D(q')$   only holds in the case of a constant function
$D(q)\equiv D$.
For each
ratio $\lambda$ there exist maximum and minimum values for the
box-probabilities. Now raising the box-probabilities to positive powers
$q\to\infty$ means that in the sum $\sum_i P_i^q$  only the
maximum values  contribute significantly
yielding the dimension $D_{\infty}>0$
by
\equ{11.18}
{
	P_{\rm max}(L_b)\propto \lambda^{D_{\infty}}\, .
}
A similar consideration for $q\to -\infty$ shows that  the minimum
values of the box-probabilities dominate in the sum $\sum_i P_i^q$
yielding the dimension $D_{-\infty}$ by
\equ{11.19}
{
	P_{\rm min}(L_b)\propto \lambda^{D_{-\infty}}\, .
}
Thus, the $\tau(q)$ function has  asymptotic slopes given by $D_{\pm
\infty}$ which shows that the  limits of $D(q)$ for $q\to\pm\infty$
exist and are just given by $D_{\pm
\infty}$. Notice  that we referred to finite ratios
$\lambda$ guaranteeing the existence of maximum and minimum values.
Extrapolating from
small but finite values of $\lambda$ to the thermodynamic limit
$\lambda\to 0$ becomes possible due to scaling,
 i.e.~the absence of a
length scale in the system.
To complete the proof we have to show that $\tau(q)$ has negative
curvature.
Due to the  smoothness assumption
\equ{11.20}
{
	\alpha(q):=\frac{d  \tau(q)}{d  q}\, ,\;
	\alpha'(q):=\frac{d  \alpha(q)}{d  q}
}
exist and are continous functions of $q$.
 The definition  of $\tau(q)$
yields
\equa{11.20a}
{
	\alpha(q) & = &
	\frac{1}{Z(q,\lambda)}\sum_i\LK\frac{\ln P_i}{\ln \lambda}\RK
	P_i^q > 0\nonumber \\
	\alpha'(q) & = &
	\LK
	\frac{\sum_i \LK\frac{\ln P_i}{\ln \lambda}\RK^2P_i^q}{Z(q,\lambda)}
	- \alpha^2(q)\RK\cdot\ln\lambda\nonumber \\
	& = &  \ln\lambda\cdot \frac{\sum_i \LK\frac{\ln P_i}
	{\ln \lambda}-\alpha(q)\RK^2P_i^q}{Z(q,\lambda)} \leq 0\,
}
where
\equ{zq}
{
	Z(q,\lambda):= \sum_i P_i^q
}
 is
called the {\em partition sum} of the multifractal. The equality in the
relation
$\alpha'(q)\leq 0$ only holds iff  $\alpha(q)\equiv
D$, i.e. in the single-fractal situation.
 From Eq.~(\ref{11.20}) one can see that
 the objects characterizing
multifractals are constructed in a similar way to thermodynamic
quantities
where the partition sum $Z(q,\lambda)$ is replaced by the canonical
partition sum $Z(\beta, V)$  in  Boltzmann statistics.
Indeed, one can write down for each of the quantities like $q, \ln
\lambda, \tau(q)\ldots$ its thermodynamic counterpart (see e.g. Ref.~${41}$).
Keeping this analogy between multifractals and thermodynamics in mind
one can imagine that a violation of the smoothness condition for
$\tau(q)$
could be viewed as an intrinsic phase transition in each multifractal
wave function. Such kind of phase transition (which has  not been observed
in systems showing LD transitions up to now) should not be    confused
 with the LD transition itself.

\subsection{Distributions}\label{subdis}
\noindent
The $\tau(q)$ function describes the scaling behavior of moments,
 i.e. of $\BRA P^q\KET$.
We wish to describe the whole
distribution function
$\Pi(P;\lambda)$ that corresponds to these moments.
For a function $F(P;\lambda)$ the corresponding average
value reads
\equ{11.21}
{
	\BRA F(P;\lambda)\KET_L = \int\limits_0^1 d  P \,
	\Pi (P;\lambda) F(P)\, .
}
Since we are interested in the scaling behavior of the distribution
function itself we change the variable $P$ to
$\alpha$ by defining
\equa{11.22}
{
	\alpha& := & \frac{\ln P}{\ln \lambda}\nonumber \\
	\tilde{\Pi} (\alpha;\lambda) d  \alpha & := & \Pi (P;\lambda)
	d  P\, .
}
Now,  average values can be calculated with the help of
$\tilde{\Pi}(\alpha;\lambda)$ via
\equ{11.23}
{
	\BRA F(P;\lambda)\KET_L = \int\limits_0^\infty d  \alpha \,
	\tilde{\Pi} (\alpha ;\lambda) F(\lambda^{\alpha})\, .
}
Absence of length scales forces the distribution
function $\tilde{\Pi}$ to show power-law scaling with respect to
$\lambda$.
Since, in the thermodynamic limit $\lambda\to 0$, the values
$\tau(q)$ are uniquely defined one can apply the method of steepest
decent
 when calculating the average for $F=P^q$.
This leads to the following conclusion
\begin{itemize}
\item
There exists a positive function,  the
$f(\alpha)$-spectrum of the multifractal, characterizing the
scaling  behavior of the distribution function $\tilde{\Pi}$, and
 is related to $\tau(q)$ by a Legendre transformation.
\equa{11.24}
{
	\tilde{\Pi} (\alpha ;\lambda) & \propto & \lambda^{-f(\alpha)
	+D}\nonumber \\
	f(\alpha(q)) & = & \alpha(q)\cdot q -\tau (q)\, .
}
Here the function $\alpha(q)$ is the same as introduced in Eq.~(\ref{11.20}).
\end{itemize}
Using the analogy between multifractality and thermodynamics one can
check that $f(\alpha)$ corresponds to the microcanonical statistics
and resembles the entropy  as a function of energy. Let's focus
on the meaning of the function $f(\alpha)$  in the context of
multifractality.  {\em This function
describes the
scaling behavior of the whole distribution function of
box-probabilities
in the absence of length scales}. In the single-fractal case
the function $f(\alpha)$ shrinks to the point $(D,D)$ in an \hbox{$(\alpha,
f(\alpha))$} diagram. Consequently, the distribution function
$\tilde{\Pi}$ is singular, and the distribution function with respect
to the variable $P$, $\Pi(P; \lambda)$, is a narrow distribution on
all length scales. In the single-fractal case
$\Pi(P;\lambda)$ is typically a gaussian distribution and can be
characterized by a few  cumulants.

In the multifractal  case there appears a new and unexpected
behavior of the distribution function $\Pi(P;\lambda)$  which
we want to describe in the following. Due to general properties of Legendre
transformations we have
\begin{itemize}
\item $f(\alpha)$ is a positive, single-humped  function of negative
curvature
on a
finite intervall of $\alpha$ values:
\equ{11.25}
{
	D_{\infty} < \alpha < D_{-\infty}\, ,\; 0\leq f(\alpha)\leq
	D\, , \; f(\alpha_0)=D \,
}
where $\alpha_0=\alpha(q=0)> D$.
\item $f(\alpha)$ terminates at the points $(D_{\pm\infty}, 0)$ with
infinite slopes and has slope $f'(\alpha)=1$  at the point
$(\alpha(1),\alpha(1))$ where $\alpha$, as well as $f(\alpha)$, equal $D(1)$.
\end{itemize}

These statements can most easily  derived  by writing
the functions $\tau(q)$,
$\alpha(q)$ and $\tilde{f}(q)$, defined by
$f(\alpha(q))=:\tilde{f}(q)$,
  as$^{42}$
\equa{11.26}
{
	\tau(q) & = & \frac{1}{\ln \lambda}\ln Z(q,\lambda)\nonumber\\
	\alpha(q) & = & \frac{1}{\ln \lambda}
	\sum_i \mu_i \cdot \ln P_i \nonumber \\
	\tilde{f}(q) & = & \frac{1}{\ln \lambda}
	 \sum_i \mu_i \cdot \ln \mu_i \, .
}
Here the $q$-dependent normalized quantity
\equ{11.27}
{
	\mu_i(q,L_b):= \frac{P_i^q(L_b)}{\sum_i P_i^q(L_b)}
}
is a generalization of the original box-probability $P_i$ and is
called the {\em $q$-microscope} since it
increases
the large box-probabilities for positive values of $q$ and increases the
small box-probabilities for negative values of $q$, respectively.
The second of these equations (\ref{11.26}) is only a reparametrization
of
Eq.~(\ref{11.20}) and by Legendre transforming $\tau(q)$ one can
easily check that $\tilde{f}(q)$ is correctly described by the third
equation
of (\ref{11.26}).
We mention that numerical calculations of $f(\alpha)$-spectra are of
higher precision when using Eqs.~(\ref{11.26}) compared to numerical
Legendre
transformation of $\tau(q)$.

Let us summarize the instruments for  describing the
multifractal scaling behavior of normalized box-probabilities: the
 $\tau(q)$ function describes the scaling behavior of moments and
$f(\alpha)$ describes the scaling behavior of the corresponding
distribution
function. Both of them  are related by  Legendre
transformation. The functions $D(q)$ and $\alpha(q)$ can also serve to
describe the multifractal nature of box-probabilities but have, for our
purposes, less appealing interpretations.

Motivated by the single-humped shape
of the $f(\alpha)$-spectrum we try to give a physical interpretation
of multifractality, at least in the context of the LD transition.
It is illuminating to approximate the $f(\alpha)$-spectrum  by a parabola
and to see to which kind of distribution function the {\em
parabolic approximation} (PA)
corresponds.
The PA, which fulfills most of the desired
constraints,
is determined by one parameter, $\alpha_0$, besides the geometric
fractal dimension $D$,$^{20,37}$ i.e.
\equ{11.28}
{
	f(\alpha)= D- \frac{\LK \alpha -\alpha_0\RK^2}{4\LK\alpha_0
	-D\RK}\, .
}
This corresponds to a log-normal distribution of the box-probabilities
\equ{11.29}
{
	\tilde{\Pi}(\alpha ; \lambda) \propto
	\exp\LB\frac{\LK \alpha -\alpha_0\RK^2}{4\LK\alpha_0
	-D\RK}\cdot\ln\lambda\RB\, .
}
A log-normal distribution is the prototype of a broad distribution
which cannot be characterized by a few
cumulants. For example, the $n$-th cumulant grows exponentially with
$n$, $\propto  e^{n^2}$.
Scaling requires that the distribution is broad on all
length scales and with the normalization condition on the
box-probabilities
the form given by Eq.~(\ref{11.29}) can be viewed as the paradigm of a
distribution function which is broad on all length scales.
The physical interpretation of such a behavior is the following.

In phase coherent
disordered conductors
the actual value of a local box-quantity, like $P_i(L_b)$, depends on a large
number of conditions {\em simultaneuosly}. This
 can be simulated by writing $P_i(L_b)$ as a
product of a large number of independent random factors,
\equ{11.30}
{
	P_i(L_b)= p_0\cdot p_1\cdot p_2 \cdot \ldots\, ,
}
and the central-limit-theorem tells that one can expect $P_i(L_b)$ to
be log-normal distributed. The scaling condition (absence of length
scales) requires the distribution function to exhibit power law
behavior. Thus,
 we conclude
\equ{11.31}
{
	\tilde{\Pi}(\frac{\ln P}{\ln \lambda};\lambda) \propto
	\lambda^{D-f(\ln P/\ln \lambda)}
}
with $f(\alpha)$ being almost parabolic.
For completeness we list the PA for  $D(q)$ and $\alpha(q)$
\equ{padq}
{
	D(q)=D -q\LK \alpha_0 -D\RK\; ,\;\;
\alpha(q)=\alpha_0 -2q\LK\alpha_0-D\RK\, .
}
However, the parabolic approximation can never be exact
since it violates some of the constraints we have derived for
$f(\alpha)$; mainly positivity and the boundary conditions at
$(D_{\pm\infty},0)$. PA can only serve as a good approximation in the
vicinity of the most probable value $\alpha_0$ for the scaling
exponents $\alpha$. The breakdown of the PA can be indicated by either
 the values $\alpha^{\pm}=\alpha_0\pm2\sqrt{D(\alpha_0-D)}$
(where $f(\alpha)$ vanishes in the PA), or by
$q_{+}=\frac{\alpha_0}{2(\alpha_0 -D)}$ (where $\tau(q)$ looses
monotonicity in the PA). $\alpha_{\pm}$
give rough estimates of $D_{\mp\infty}$.

 The most probable value of the box-probability
corresponding to $\alpha_0$ is given by  the {\em typical value}
of $P(L_b)$,
\equ{11.32}
{
	P_{\rm typ}(L_b):= \exp\LB \BRA \ln P(L_b)\KET_L \RB \propto
	\lambda^{\alpha_0}\, .
}
which is a geometric type of mean value.

Finally, we present  results of the multifractal analysis
for  wave functions in the critical regime of a finite  quantum Hall
system.$^{20}$
In Fig.~\ref{Fig.11.5} the squared amplitude
of a wave function is
 shown where increasing darkness correponds to higher probability to
find an electron at the corresponding point in the system.
\fig{8}{Squared amplitude of a multifractal wave function in the
critical regime of a quantum Hall system. Increasing
darkness correponds to higher probability of
finding an electron (on a logarithmic scale).}{Fig.11.5}
The
picture  demonstrates the self-similar structure of the wave
functions
and the tremendeous amount of fluctuation of local probabilities, even
 within one
given state. In Fig.~\ref{Fig.11.6}
 $f(\alpha)$ spectrum  for
this wave function is graphically depicted, showing
that
the box probabilities are almost log-normal distributed and show power
law behavior.
The most interesting values  of scaling exponents are
\equ{11.33}
{
	\alpha_0 = 2.3\pm 0.07 \, ,\; D_{\infty} = 0.95 \pm 0.1 \, , \;
	D_{-\infty}= 3.7 \pm 0.1 \, .
}
The fractal dimension is, of course, $D=2$.
\fig{8}{The $f(\alpha )$ spectrum of a multifractal wave
function at the center of the lowest Landau level.}{Fig.11.6}

Striking is  the observation that the $f(\alpha)$
function seems to be {\em universal} for all of the critical states. This
can be seen in Fig.~\ref{Fig.11.7} where the $f(\alpha)$-spectrum of one
state is shown together with the parabolic approximations of this and
two distinct states choosen in the critical regime. The coincidence is
remarkable.
\fig{8}{Comparison of $f(\alpha)$-spectra for three different
  wave functions in the critical regime of a quantum Hall system.
The dots correspond
to the $f(\alpha)$-spectrum  and other curves
correspond to the parabolic approximations for the $f(\alpha)$-spectra
of this and two distinct states.}{Fig.11.7}
 Calculations of $f(\alpha)$-spectra,
carried out by Pook$^{20}$ for the same
model system, but with
different realizations of the random potential and different system sizes,
 confirm this result. Further evidence was provided by calculations of
Huckestein et al.$^{22,43}$
for a tight-binding version of the quantum
Hall system which give the same values for the quantities $\alpha_0$,
$D_{\pm\infty}$  within the error bars. This means
that the LD transition of quantum Hall systems (at least for the lowest
Landau level where all of the numerics has been performed)
 is uniquely characterized by those critical numbers.

\subsection{Inverse Participation Numbers}\label{subinv}
\noindent
In this subsection we describe how disorder-averaged   Green functions
are involved  in the multifractal analysis via
 inverse participation numbers.
Thereby  two applications  will be adressed: the first
 concerns the calculation of the critical exponent $\nu$
of the correlation (localization) length
in a quantum Hall sysytem. The second
 concerns the anomalous diffusion found by Chalker et al.$^{30,44}$
 as mentioned in Sec.~$2.4$.

The role of fractality for the LD transition was already observed by
Aoki$^{45}$ in 1983 by looking at the inverse participation
number defined by
\equ{11.34}
{
	{\cal{P}} = \int\limits_{\Omega} {d }^d r\,
	|\psi (\vec{r})|^4 \,
}
where $\Omega$ denotes a $d$-dimensional
system with linear dimension $L$.
He gave an argument that the wave function $\psi (\vec{r})$
of electrons at the LD transition point shows fractal structure.
If the wave function is uniformly
distributed, ${\cal{P}}\propto L^{-d}$,
the participation ratio $p=({\cal{P}}L^d)^{-1}$ is constant.
In the localized regime ${\cal{P}}\approx \xi^{-d}$ and $p$ vanishes
in the thermodynamic limit.
At the transition point where the wave function is extended
the participation ratio still
vanishes in the thermodynamic limit.
Consequently, ${\cal{P}}$ scales with a power $d^{*}<d$.
By a box-counting method Aoki$^{46}$
 calculated the fractal dimension of wave
functions of a QHS at the transition point to be approximately $1.5$.
Although his method was not free of systematic errors, plots of the electron
density $|\psi (\vec{r})|^2$ indicated self-similarity.

Castellani et al.$^{19}$
 considered the ``$d=2+\epsilon$'' expansion
of Wegner's non linear $\sigma$-model$^{47}$ for
the generalized inverse participation numbers
\equ{11.35}
{
	{\cal{P}}^{\lbrack q\rbrack} = \int\limits_{\Omega}
	{d }^d r |\psi (\vec{r})|^{2q}
	\propto L^{-\tau^{*}(q)}
}
and concluded that the wave functions at the transition point
show multifractal behavior since $\tau^{*}(q)$ is not equal to
$d(q-1)$ but defines a set of fractal dimensions $d^{*}(q)$:
\equ{11.36}
{
	\tau^{*} (q)=d^{*}(q)\left( q-1\right) \, .
}
To show the relation between $\tau^{*}(q)$ and $\tau(q)$ we consider
 ${\cal{P}}^{\lbrack q\rbrack }$ calculated in a tight binding
 model with lattice
constant $b$,
\equ{11.37}
{
	{\cal{P}}^{\lbrack q\rbrack} = \sum_i |\psi (i)|^{2q} =
	Z(q,\lambda=b/L)
						\, .
}
If both quantities $b$ and $L$ are in a regime of self-similarity
the identity of $\tau^{*}(q)$ and $\tau(q)$ follows immediately.
However, they are different in the continuum limit ($b\to 0$)
for fixed $L$ since self-similarity breaks down once $b$ is smaller
than the microscopic length scales $l$.
On the other hand, in the continuum limit, we can consider
$Z(q,\lambda=b/L)$
to be an approximation for ${\cal{P}}^{\lbrack q\rbrack}$
which becomes more and more accurate as $L$ goes to infinity.
If self-similarity is preserved up to length scales $L\to\infty$,
$\tau^{*}(q) \to \tau(q)$  can be concluded.
The assumption of self-similarity for
$L\to\infty$ is fulfilled at the critical point of the LD transition
where the only relevant length scale, the localization length $\xi$, diverges.
Thus, at the critical point of the LD transition
\equ{11.38}
{
	\tau^{*}(q)=\tau(q)\, .
}
On the other hand the correlation length $\xi_c$  is defined as an
upper bound for  power law behavior and $d^{*}(q)$ can be determined
from the scaling
behavior of ${\cal{P}}^{\lbrack q\rbrack}$ with $\xi_c$:
\equ{11.39}
{
	{\cal{P}}^{\lbrack q\rbrack}\propto \xi_c^{-\tau^*(q)}
	\, .
}
The critical exponent $\nu$ of the correlation length $\xi_c$ is defined
by
\equ{11.40}
{
	\xi_c
	 \propto \Delta^{-\nu}
}
where $\Delta$ is the difference of a critical observable
(e.g.~the energy) from it's
critical value.
In analogy to this, Wegner$^{47}$
 introduced critical exponents $\pi(q)$
for the generalized inverse participation numbers
\equ{11.41}
{
	{\cal{P}}^{\lbrack q\rbrack}
	\propto \Delta^{\pi(q)} \, .
}
Thus, we conclude a scaling relation between multifractal scaling
numbers $\tau(q)$ and critical exponents $\pi(q)$ and $\nu$$^{17,48}$,
\equ{11.42}
{
	\pi(q)=\nu\cdot \tau(q) 		\, ,
}
and $ \pi(q)$ depends on $q$ in a non-trivial way if the wave functions
are multifractals.  Still, Eq.~(\ref{11.42})
doesn't yield $\nu$ as a function of only
multifractal scaling numbers.
Nevertheless,
 we can see what was overlooked by Hikami$^{49}$ when calculating
the critical exponent $\nu$
for a QHS (restricted to  one Landau level)
to be $1.9\pm 0.2$. What he did calculate was the exponent $\pi(2)$.
He found $\pi(2)=3.8\pm 0.4$ The result for $\nu$ given above follows if
${\cal{P}}^{\lbrack 2\rbrack}$ is assumed to scale with the localization
length $\xi$ as $\xi^{-2}$,
ignoring  that multifractality leads to the anomalous
scaling behavior ${\cal{P}}^{\lbrack 2\rbrack} \propto \xi^{-D(2)}$.
Taking multifractality into account, $D(2)=1.62\pm 0.02$,$^{50}$
we get
\equ{11.43}
{
	\nu = 2.4\pm 0.3 \, .
}
This is in agreement with the high accuracy result, $\nu =2.34\pm0.04$, of
Huckestein and Kramer$^{51}$
obtained by a finite size scaling technique.

As a further application to the QHE we show that multifractality relates
the exponent $\eta$ of anomalous diffusion to $D(2)$.
We have already anticipated that the spectrum of multifractal dimensions
has universal features for states in the vicinity of the LD
transition since we omitted to distinguish between inverse participation
numbers of individual states and their ensemble average.
In fact, as already discussed,
 different wave
functions of several systems show, in the critical regime,
the same $f(\alpha)$-spectra within the error bars.

The observation of anomalous diffusion (see Sec.~$2.4$)
 means that the averaged
two-particle Green function
\equ{11.44}
{
	\overline{|G^{+}(r,E)|^2}=\overline{|\langle\vec{r}|
	(E-H+i\epsilon)^{-1}|\vec{0}\rangle|^2}
}
behaves as
\equ{11.45}
{
	\overline{|G^{+}(r,E)|^2}\propto r^{-\eta+2-d}\,
}
This power law, with a value  $\eta =0.38\pm 0.04$,
was found$^{30}$ in quantum Hall systems (d=2)
 for length scales $r$
in the regime of multifractality, i. e. $l\ll r\ll \xi$.
Replace the ensemble average of the inverse participation ratio
(\ref{11.36})
of wave functions $\psi_{\varepsilon_\alpha}(i)$
with
\equ{11.46}
{
	{\cal P}^{[q]}(E):=
	{\overline{\sum_{\alpha}\sum_i |\psi_{\varepsilon_\alpha}(i)|^{2q}
	\cdot \delta (E-\varepsilon_\alpha)}}/{\overline{\sum_\alpha
	 \delta(E-\varepsilon_\alpha)}}
}
where  $\psi_{\varepsilon_\alpha}(i)$ corresponds to
 energy eigenvalue $\varepsilon_\alpha$.
 	The ensemble averaged inverse participation number
${\cal{P}}(E)$ is now defined with respect to  energy. It
can
be calculated from the Green function$^{47}$
\equ{11.47}
{
	{\cal{P}}(E) =\lim\limits_{\epsilon\to +0}
	\frac{\epsilon}{\pi \rho(E)}\overline{|G^{+}(0,E)|^2}
}
where \hbox{$\rho(E)=\lim\limits_{\epsilon\to +0}
	\frac{\epsilon}{\pi}\int\limits_{\Omega}
	 {d }^d r\, \overline{|G^{+}(r,E)|^2}$}
is the density of states. The expected scale independence of this
quantity
is  excellently confirmed
by numerical results$^{20}$.
Thus, with the ansatz
\equ{11.48}
{
	\overline{|G^{+}(r,E)|^2} =\gamma(E,L)r^{-z}
}
one concludes
\equ{11.49a}
{
	{\cal{P}}(E)\propto \gamma(E,L)\propto L^{z-d}\, ,
}
and consequently
\equ{11.49}
{
	\eta + d -2=z=d-D(2)\, .
}
More precisely,
in the case of multifractal behavior one expects
the Green function to have a spectrum of exponents and
the one occurring in Eq.~(\ref{11.49}) should be interpreted as the dominating
one.
In the regime of multifractality, besides the distance $r$,
 a second
length scale, the system size $L$, appears in the correlation function
$\overline{|G^{+}|^2}$  (Eq.~(\ref{11.48})). This was
anticipated in Sec.~$2.4$ when discussing the difference between
the distance exponents $z$ and $\tilde{z}$. Using the terminology of
Sec.~$2.2$ we identify the exponents $y$ and $\tilde{z}$ to be
\equ{11.49aa}
{
	y=-z=D(2)-d \;\; ,\;\;\; \tilde{z}=y+z=0\, .
}
In  a single-fractal
situation, $z=0$ and $y=\tilde{z}$.
Thus, in systems without  multifractal structure there is no need for
introducing
two exponents when discussing correlation functions in finite systems.
The numerical value$^{50}$
$D(2)=d^{*}(2)=1.62\pm 0.02$ is in
accordance with our considerations,
 and
 anomalous diffusion can be interpreted as being a direct consequence of the
multifractal nature of wave functions at the critical point of the LD
transition. This was already conjectured in Ref.~$30$.
 There is no contradiction between a finite value of
$\eta$, as interpreted via the distance exponent $z$ instead of
$\tilde{z}$,
and the vanishing of the exponent $\beta$ corresponding to the formal
order parameter of the LD transition.
The physical reason for the phenomena of anomalous diffusion and of
multifractality  is the quantum
coherence of  highly disordered systems.

\section{Genralized Multifractal Analysis}\label{secgen}
\noindent
So far, only measures, i.e.   box-probabilities
$P$, have been considered.
To get into contact with usual
critical exponents we introduce a generalized multifractal
analysis\index{generalized multifractal analysis}
for non-normalized observables.$^{20}$ A related
approach in the context of field theory
was put forward by A. Ludwig.$^{24}$

Consider a physical observable $Q$ which can be defined
for systems of any linear extension  $L$. For simplicity, we assume the
system to be a box of volume $(L_b)^d$ where $d$ is the euklidean dimension
of the system. Thus, for any of such boxes we have a {\em
box-observable} $Q_i(L_b)$ similar to the
box-probability in the foregoing sections.
 Examples are the magnetic moment or the
two-terminal conductance
of a $d$-dimensional cube with volume $(L_b)^d$.

We are going to study the statistical and  scaling
properties of box-observables. We restrict to {\em positive}
values for $Q$,
\equ{positiv}
{
	Q(L_b)\geq 0\, ,
} to avoid cancellations in averaging procedures and to
apply the results of the foregoing sections.
At  first glance,
this seems to be a strong
restriction. However, still a large class of observables can be
studied which is specified by the following consideration.
Let a physical observable $A(L_b)$, measured on a box, be an element
of
some observable algebra, then the scaling behavior of $A(L_b)$
may show up in a positive scalar factor $Q$,
\equa{11.50}
{
	A(s\cdot x) & = & s^\kappa\cdot A(x)\nonumber \\
	A(x) & = & Q(x)\cdot A'(x)\nonumber\\
	Q(x) & \propto & x^\kappa\,
}
where the algebra element $A'$ does not show significant scaling
behavior, or it's average might even vanish due to symmetry.
The following generalized multifractal analysis applies to such observables.
With this restriction in mind we proceed as follows.

\subsection{Normalization Exponents}\label{subnor}
\noindent
To study the statistics of box-observables
one takes a large number ${\tilde{N}}$ of equivalently prepared systems
and records the values of the box observable. This leads to the set
$\LB Q_i(L_b)\RB_{i=1,\ldots ,{\tilde{N}}}$. The mean value of a function of
$Q$
is then calculated as
\equ{mean}
{
	\BRA F(Q(L_b))\KET_{\tilde{N}}
:=\frac{1}{{\tilde{N}}}\sum_i^{\tilde{N}}  F(Q_i(L_b))\, .
}
A corresponding smooth
distribution function $\Pi (Q,L_b,{\tilde{N}})$ is assumed to describe
average values (provided ${\tilde{N}}$ is large enough),
\equ{mean2}
{
	\BRA F(Q(L_b))\KET_{\tilde{N}} = \int\limits_0^{\infty} dQ \, F(Q) \Pi
(Q,L_b,{\tilde{N}})\, .
}
 In the absence of intrinsic length scales, i.e.  compared with $L_b$
intrinsic length scales are either much less or much larger than
$L_b$,
one expects power law scaling to occur in averages.
To make contact to the multifractal analysis of measures
at an early state of the investigations let us introduce
a {\em symbolic} system size $L$ by reparametrizing the number ${\tilde{N}}$ as
\equ{syml}
{
	L:=\LK L_b^d {\tilde{N}}\RK^{1/d}
}
Let $N\leq {\tilde{N}}$ be the number of non-vanishing values for $Q_i(L_b)$,
define the {\em normalization} of $Q$ by
\equ{norm}
{
	||Q(L_b,L)||:= \sum_i^N Q_i(L_b)
}
and, by keeping $L$ fixed in the averaging procedure,
identify the {\em normalization exponent} $X^{[Q]}$
in the regime of power law scaling via
\equ{norm2}
{
	||Q(L_b,L)||\propto L_b^{X^{[Q]}}\, .
}
$X^{[Q]}$  can be any real number.
The crucial step is the construction of a {\em normalized
box-observable}
$P([Q],L_b,L)$ associated with $Q$:
 \equ{11.51}
{
	P_i ([Q],L_b,L) =\frac{Q_i(L_b )}
	{||Q(L_b,L)||}\; ,\;\; \sum_i^N P_i([Q],L_b,L)\equiv 1\, .
}
 $P_i([Q],L_b,L)$ is normalized with respect to the symbolic
system of linear size $L$ but, in general, not additive,
\equ{11.52}
{
	P([Q],L) \neq \sum_i^{N} P_i([Q],L_b,L)\, .
}
In the usual multifractal analysis one imagines the box-observables to
be measures for which the normalization condition is very natural.
Nevertheless,
 when proving the general features of the functions
$\tau(q)$ and $f(\alpha)$  the  normalization
condition was essential, but  no use was made of additivity.
By fixing $L$
 we are able to attach to the positive, normalized box-observable
$P([Q],L_b,L)$ the multifractal analysis by defining $\tau^{[Q]}(q)$,
$\alpha^{[Q]}(q)$ and $f^{[Q]}(\alpha^{[Q]})$ with the help of the
partition sum,
\equ{11.53}
{
	Z([Q]; q,\lambda):= \sum_i^{N} P_i^q([Q],L_b,L)\, ,
}
in direct analogy to Eqs.~(\ref{11.26}). For these functions all the general
features  explained in Sec.~$3$
 follow if the ratio $\lambda=L_b/L$ is
choosen small enough, and the observables $Q_i(L_b)$ are in the scaling
regime.  The  results of Sec.~$3$
can be  translated to the following statements
\begin{itemize}
\item
	For positive box-observables $Q(L_b)$ the  scaling behavior of
its distribution function can be described by the corresponding
normalization exponent $X^{[Q]}$ and a single-humped  function
$f^{[Q]}(\alpha^{[Q]})$
\equ{11.56}
{
	\tilde{\Pi}\LK\frac{\ln Q}{\ln \lambda},L_b\RK\propto
	L_b^{D-h\LK \frac{\ln Q}{\ln \lambda}\RK}
\; ,\;\; h\LK \frac{\ln Q}{\ln \lambda}\RK
=f^{[Q]}\LK\alpha^{[Q]}\RK
}
 with
\equ{with}
{
		\frac{\ln Q}{\ln \lambda} = \alpha^{[Q]} +X^{[Q]} \, .
}
\item
    The scaling relations for the moments and the typical value,
\equ{typv}
{
	Q_{\rm typ}(L_b):= \exp\LB\BRA \ln Q(L_b)\KET\RB\, ,
}
 read
\equa{11.57}
{
	\BRA P^q([Q],L_b,L) \KET & \propto &
	L_b^{D+\tau^{[Q]}(q)}\nonumber\\
	\BRA Q^q(L_b) \KET & \propto & L_b^{x(q)}\; , \;\;
x(q)=D+\tau^{[Q]}(q) +q\cdot X^{[Q]}
	\nonumber\\
	Q_{\rm typ}(L_b) & \propto & L_b^{\alpha^{[Q]}_0 + X^{[Q]}}
}
 The statements of Sec.~$3$
about properties of $\tau$, $\alpha$, $f$
such as monotonicity, curvature, Legendre transform, etc.
 do also apply for $\tau^{[Q]}, \alpha^{[Q]}$ and
$f^{[Q]}$. In general one expects  $D=d$, unless a systematic
scaling dependence of
${\tilde{N}}(N)$ appears.
\end{itemize}

\subsection{Local observables}\label{subloc}
\noindent
In case that $Q$ is a local additive observable, i.e. for each subdivision
of a starting system of linear size $L$
into $N$ boxes of linear size $L_b$,
$Q(L_b)$ is an observable and additivity
holds,
\equ{additi}
{
	Q(L)=\sum_i^N Q_i(L_b),
}
the generalized multifractal analysis can also be applied to the
investigation
of scaling properties with respect to the system size $L$ which is
then
a {\em true} system size. We focus on the situation where $L_b$ and
$L$ are in a regime of power law scaling and vary both lengths
independently.
We thus have to be aware of an additional normalization exponent $Y^{[Q]}$
with
respect to $L$ which is, in general, different from $X^{[Q]}$,
\equ{norm4}
{
	||Q(L_b,L)||\propto L_b^{X^{[Q]}}L^{-Y^{[Q]}}\, .
}
The box-probabilities studied in Sec.~$3$
fall, of course, into the class of local observables with
trivial normalization exponents $X=Y=0$.
For local observables one finds, besides the scaling relations of
Eqs.~(\ref{11.56}, \ref{11.57}) with respect to $L_b$,
 (by varying $L$ and taking the normalization exponent $Y^{[Q]}$ of
Eq.~(\ref{norm4}) into account)
\equ{addob}
{
 \BRA Q^q(L_b) \KET_L \propto  L_b^{x(q)}L^{-y(q)}\; ,\;\;
	Q_{\rm typ}(L_b,L)  \propto  L_b^{\alpha^{[Q]}_0 +
	X^{[Q]}}L^{-\LK\alpha^{[Q]}_0 + Y^{[Q]}\RK}
}
with
\equ{addob2}
{
	x(q)=D +\tau^{[Q]}(q)+qX^{[Q]}\; ,\;\;
y(q)=D+\tau^{[Q]}(q)+qY^{[Q]}\, .
}
As an illustration let's discuss two examples. A trivial example
is the box-observable $Q:=PL_b^{X}L^{-Y}$ where $P$ is a box-probability
with multifractal functions $\tau$, $\alpha$ and $f$ and $X,Y$ are
real numbers.
Then,  $\tau^{[Q]}=\tau $, $\alpha^{[Q]}=\alpha $, $f^{[Q]}=f$, and
$X^{[Q]}=X$, $Y^{[Q]}=Y$.
As another example consider the box-observable  $Q=P^m$ where $P$ is a
box-probability  and $m$   some real number.
 All the relevant
scaling information is already contained in the $\tau(q)$ function of $P$.
Thus,
it is easy to check that
\equa{11.58a}
{
	X^{[P^m]}= Y^{[P^m]} & = & \tau(m)\nonumber\\
	\alpha^{[P^m]}(q) & = & m\cdot\alpha(q\cdot m) -
	X^{[P^m]}\nonumber\\
	f^{[P^m]}(\alpha^{[P^m]}(q)) & = & (qm)\cdot \alpha(qm)
	-\tau(qm)\, .
}
The last of these three equations yields, due to the Legendre
transform property of $f(\alpha)$ with respect to $\tau(q)$,
\equ{fast}
{
	f^{[Q]}(\alpha^{[Q]})=f(\alpha(qm))\, .
}
 It is tempting to expect
from this examples that $f(\alpha)$-spectra of different
observables, defined for one ensemble of equivalently prepared systems,
are either identical, if only the normalization
is distinct, or are related in a simple manner, if powers of
elementary
observables are involved.

\subsection{Typical Scaling Variables}\label{subtyp}
\noindent
In the vicinity of a phase transition one expects a universal
behavior of the scaling exponents. They may depend on
general properties of the systems dynamic, however not on
microscopic details.
We have already observed that different wave functions in the LD
transition regime showed the same $f(\alpha)$-spectrum. A further
indication
for universality of $f(\alpha)$-spectra was provided by the
observation of Huckestein and Schweitzer$^{43}$ that the
local equilibrium current density as well as the local magnetization$^{52}$
 show the same
$f(\alpha)$-spectrum
as the wave functions in a quantum Hall system. However, universality
of scaling variables in the LD transition (which are non equilibrium
properties)
is  even more interesting.

Recall the phenomenological scaling theory of Abrahams et al.$^{5}$
 for the LD transition (see Sec.~$2.4$) where
the conductance $g$ is assumed to be a scaling variable on all length scales.
As we already mentioned in Sec.~$2.4$
the conductance of mesoscopic
systems has a broad distribution which cannot be described
in terms of the mean value of $g$. Consequently, one has to consider
the distribution function  $\Pi(g,L)$ which can be characterized by
relevant parameters $g_{\rm rel}$ for which $\beta$-functions exist.
If  $g$ is choosen for the $f(\alpha)$ analysis
a candidate for ${g_{\rm rel}}$ is obviously given by the typical value
\hbox{$g_{\rm typ}(L)=\exp\LB\BRA \ln g(L)\KET\RB$}
\equ{11.61}
{
	g_{\rm typ}(L)- g_{\rm typ}^*
	\propto L^{\alpha^{[g]}_0 +X^{[g]}}
}
and by comparing Eqs.~(\ref{8.7.3},\ref{11.57}) one concludes
\equ{11.62}
{
	1/\nu = \alpha^{[g]}_0 + X^{[g]}\, .
}
The universality of $\nu$  suggests  the $f(\alpha)$-spectra of
normalized scaling observables being universal, too.
Notice that Eq.~(\ref{11.62}) provides  a method to calculate
the critical exponent $\nu$ with only the help of multifractal
critical numbers. This gives the possibility for calculating $\nu$ by
a different method than the finite size scaling method.$^{14,15}$
 In addition, if both methods
coincide
the multifractal analysis  demonstrates
that typical observables are relevant for scaling; a consequence of
the fluctuating self similar structure underlying multifractal
objects.

In order to check the predictions of the multifractal analysis
  Fastenrath et al.$^{23}$
calculated numerically Thouless numbers
for the QHS and applied the multifractal analysis to the data.
The Thouless numbers  are defined by the ratio of the energy
shift due to a change of boundary conditions (e.g. periodic
$\longrightarrow$ anti periodic) with respect to the mean level spacing.
They can be thought of as being a transport quantity showing the same
qualitative (and scaling) behavior of a (dimensionless)
dc conductance.$^{53}$

In Fig.~\ref{Fig.11.8} data for the Thouless numbers denoted by $g$
are shown which
are calculated for the
lowest Landau level of $6$ realizations of a
QHS with linear size $120 r_c$.
\fig{8}{Thouless numbers for $6$ realizations of a quantum Hall system
with linear size $120 r_c$ restricted to the lowest Landau level.
$\omega_c$ is the cyclotron frequency. Energy windows for data
evaluation as described in the text are indicated by vertical
lines (after Ref.~$23$).}{Fig.11.8}
One can see the critical
regime outside of which the states are localized and have zero dc
conductance.
In the critical regime there are large fluctuations and the
corresponding histogram of Fig.~\ref{Fig.11.9} demonstrates the broadness
of the distribution function.
\fig{8}{Histogram of
Thouless numbers calculated for the lowest Landau level  of one
quantum Hall system.}{Fig.11.9}
To apply  the multifractal analysis, Thouless numbers $g(L)$
for a large number of different systems with varying system sizes from
$30r_c$ to $120r_c$were calculated.
As the box observable $Q$ the modulus of the deviation
$Q(L):=|g(L)-g_{\rm typ}^{*}|$ was chosen where
 $g_{\rm typ}^{*}$ was
indeed size independent.   \hbox{Fig.~\ref{Fig.11.10}}  shows
that the corresponding
$f(\alpha)$-spectrum of the Thouless numbers is very similar to
that of the box-probabilities.$^{20}$
\fig{8}{$f(\alpha)$ spectra of normalized (right) and unnormalized
(left) Thouless numbers. The displacement of the extrema is just the
normalization exponent $X$. The broken line to the right is a
parabolic approximation (after Ref.~${23}$).}{Fig.11.10}
Data have been taken from a narrow window in the
critical region as indicated in \hbox{Fig.~\ref{Fig.11.8}}.
The authors  find
\equ{11.63}
{
 	\alpha^{[g]}_0 = 2.22 \pm 0.05\, , \; X^{[g]}
	= -1.75 \pm 0.05\, , \; \nu
	=2.2 \pm 0.3\, .
}
 They also calculated $\nu$ directly from
\hbox{$|\ln {g_{\rm typ}} -\ln {g_{\rm typ}}^{*}|\propto L^{1/\nu}$}
and  found
\equ{11.64}
{
	\nu =2.3\pm 0.2\, .
}
In either case they used \hbox{$g_{\rm typ}^{*}=0.27 \pm 0.03$}, determined at
the band center.
 The data for $\nu$ are  in agreement with the high precision value of
Huckestein and Kramer, obtained by finite size scaling calculations.$^{51}$

The multifractal analysis of Fastenrath et al.$^{23}$
 confirm the one-parameter scaling theory of the LD transition
 in terms of the typical conductance and demonstrate the broadness
of the conductance distribution at criticality.

\subsection{Bounds for the Correlation Length Exponent}\label{subbou}
\noindent
In this section we show how the multifractal analysis provides a lower
bound for the critical exponent $\nu$.  We refer to
the LD transition for concreteness but the results generality
is reflected by the fact that only a few assumptions such as
one-parameter scaling  are needed. To obtain upper bounds for $\nu$
more restrictive assumptions are needed.

That $\nu$
has a lower bound given by
\equ{11.65}
{
	\nu > \frac{2}{d}
}
where $d$ is the Euclidean space dimensionality is known since the
work of Chayes et al.\,.$^{54}$
A trivial upper bound for $\nu$ is given by
\equ{11.66}
{
	\nu < \infty
}
otherwise the definition of the exponent would be meaningless (e.g.~in
the Kosterlitz-Thouless transition).$^{55}$
We give arguments which improve both bounds. Our arguments rely on
the assumption that one parameter scaling holds true in the vicinity
of the LD transition\index{LD transition}
 point, on the analytic properties of $\tau(q)$
functions and on the universality hypothesis.

In a field theoretical
statistical
model for a critical phenomenon as outlined in Sec.~$2$ a scaling
operator\index{scaling operator}
$\Theta$ couples to the scaling field $t$ (in the LD transition problem $t$
corresponds to
  $|E-E_c|=:t$) in the Hamiltonian
\equ{11.67}
{
	t\cdot \Theta = t\cdot \int d^{d} r\, S(\vec{r})\, .
}
Here $S(\vec{r})$ is a local scaling operator
of the field theoretical
statistical model.
The scaling dimensions $y(n)$ corresponding to the scaling operators
\equ{11.68}
{
	\Theta^{[n]}:=\int d^{d} r\, S^n(\vec{r})
}
are defined with respect to the renormalization group
action (see e.g.~Ref.~$56$).
Renormalizability\index{renormalizability}
of the theory then means that only a finite number of these exponents
are positive (relevant) and most of them are negative (irrelevant).
Here we identify them via the finite size scaling properties of the average
\equ{11.69}
{
	\BRA \int\limits_{\Omega} d^{d} r\,
	S^n(\vec{r})\KET_L \propto L^{\kappa(n)}
}
where the statistical model is defined on a finite system $\Omega$
of linear
size $L$. For a relevant scaling operator
the integral diverges for $L\to\infty$ and consequently
the scaling fields $t^{[n]}$ in $t^{[n]}\cdot\Theta^{[n]}$
 have to be zero, i.e.~the critical point is
reached. One-parameter scaling\index{one-parameter scaling}
 means that there is exactly
one
relevant operator and thus
\equ{11.70}
{
	\kappa(n) < 0 \, \; {\rm for}\;\; n\geq 2\, .
}
If by accident or symmetry properties the expectation value in
Eq.~(\ref{11.69}) vanishes  we {\em assume}
that there exists  a corresponding positive scalar observable
$Q$ which has the same scaling behavior as $S(\vec{r})$, in the spirit
of Eq.~(\ref{11.50}).
If this assumption holds true, then the general analytic properties of
$\tau(q)$ functions  tell that
\equa{11.71}
{
	\kappa(n) &=& -\LK \tau^{[\Theta]}(n) +X^{[\Theta]}\cdot
	n\RK\nonumber\\
	\frac{d  \kappa(n)}{d  n} &=& -\LK\alpha^{[\Theta]}(n)+X^{[\Theta]}\RK
	\nonumber\\
	\frac{d^2 \kappa(n)}{d  n^2} & \geq 0 & \, .
}
Consequently the scaling dimensions
 have positive curvature.
Since the exponent $\nu$ is given by $y(1)=1/\nu$ (for consistency with
the divergence of the correlation length) we have
\equ{11.72}
{
	X^{[\Theta]} = - 1/\nu
}
In order to have $\kappa(n)$ monotonically decreasing
guaranteeing renormalizability the lower bound
\equ{11low}
{
	\nu>1/D^{[\Theta]}_{\infty}
}
has to be respected. Furthermore,
 the one-parameter scaling condition Eq.~(\ref{11.70})
requires
\equ{11.73}
{
	\nu > \frac{2}{D^{[\Theta]}(2)}\, .
}
This resembles and improves the Chayes et al. criterion Eq.~(\ref{11.65}).
The observation of universality in $f(\alpha)$ spectra for
different observables suggests
 $D^{[\Theta]}(q)=D(q)$. Relying on the universality hypothesis for
$f(\alpha)$
 the multifractal analysis of the wave
function
in the LD transition problem already allows to obtain a lower bound
for $\nu$, $\nu > 2/D(2)$.

In Sec.~$4.3$ we have found that it is the typical
conductance $g_{\rm typ}$ rather than the
average conductance $\BRA g\KET$ which serves as a one-parameter
scaling variable for the LD transition and scales like
\equ{11.75}
{
	|g_{\rm typ}(L)-g^{*}_{\rm typ}|\propto L^{1/\nu}\, .
}
 The corresponding $\beta$-function has an unstable fixed point.
The exponents $x^{[g]}(q)$ describing the scaling of moments $\BRA
|g(L)-g^{*}_{\rm typ}|^q\KET\propto L^{x^{[g]}(q)}$ are given by
\equ{11.76}
{
	x^{[g]}(q)=D+\tau^{[g]}(q)+qX^{[g]}\, .
}
The analytic properties of $x^{[g]}$ are summarized as follows.
\equa{xgvonq}
{
	 \frac{d x^{[g]}(q)}{dq}& = &
\alpha^{[g]}(q)+X^{[g]}\nonumber\\
	\frac{d^2 x^{[g]}(q)}{dq^2} & = & \frac{d \alpha^{[g]}(q)}{d
q} < 0\, .
}
Since $x^{[g]}(0)=0$ and $1/\nu=\alpha^{[g]}_0 + X^{[g]}>0$
the exponents $x^{[g]}(q)$ are positive for small positive values of
$q$.
For large enough values of $q$
they will become negative if $D^{[g]}_{\infty} + X^{[g]}<0$ which
means
that high moments of $|g(L)-g^{*}_{\rm typ}|$ may decrease when
approaching
the transition point. This is not in contradiction to
renormalizability
and does not yield a rigorous upper bound for $\nu$.
However, if at least the first moment
has positive scaling exponent, $x^{[g]}(1)=D+X^{[g]}>0$,
then an upper bound for $\nu$ can be concluded
\equ{11.78}
{
	\nu < \frac{1}{\alpha^{[g]}_0 -D}\, .
}
Note, that this upper bound relies on the additional, though intuitive,
 assumption
about a positive scaling exponent for the mean deviation $\BRA
g(L)-g^{*}_{\rm typ}\KET$.

We give three examples for which the validity of the bounds can be
checked relying on the universality of $f(\alpha)$ spectra.

\noindent 1. For the QHS the values for $\nu$, $D(q)$ and $\alpha_0$
reported above
are compatible with the bounds,
\equ{11.80}
{
	1.3 < \nu =2.3 < 3.3\, .
}
2. A similar conclusion holds
for the 2-d spin-orbit model (the calculation of $\nu=2.75$ was done by finite
size scaling methods$^{11}$  and $\alpha_0=2.175$
 was calculated by the multifractal
analysis$^{57}$),
\equ{11.80b}
{
	1.2 < \nu= 2.75 < 6
}
3. The one-loop results of
Wegner's
non linear $\sigma$-model$^{47}$ in  $d=2+\epsilon$
leads to a parabolic $f(\alpha)$-spectrum as shown in Ref.~${19}$.
 The results are
\equ{11.79}
{
	\nu=1/\epsilon\,  ,  \; \alpha_0 = d+\epsilon\, , \; D(2)=d-2\epsilon
	+\frac{1}{2}\epsilon
}
Thus, the one loop result meets the upper bound and the lower bound is
valid up to $\epsilon < 2/3$. Higher loop orders$^{47}$ lead to $f(\alpha)$
spectra which violate the condition of constant curvature if taken
seriously for arbitrary values of $\epsilon \leq 1$. This indicates  that
the higher orders are improvements only for very small values of
$\epsilon$
(as often happens in asymptotic expansions).
The applicability of the loop expansion is an
open question.

\section{Correlations in Multifractals}\label{seccor}
\noindent
In this section we ask for correlations in positive, local observables
as introduced in Sec.~$4.2$. For example, a local field
$\varphi(\vec{r})$ gives rise to local box-observables
\equ{17local}
{
	Q_i(L_b)= \int\limits_{\rm box (i)} d^d r\,
\mid \varphi(\vec{r})\mid\, .
}
We are interested in the scaling properties of
\equ{m}
{
	M^{[q]}(r,L_b,L):=\BRA Q_i^q(L_b)Q_{i+s}^q(L_b)\KET_L
}
where the average is over all pairs of boxes with fixed distance
$r=sL_b$.

\subsection{Scaling Relations}\label{subsca}
\noindent
Usually, in critical phenomena one studies correlations for infinite
system size (and $L_b$ being microscopic)
as a function of $r$ alone. Here we consider a regime,
where
both $r$ and $L$ are able to show scaling behavior. We thus start
with an ansatz
\equ{mansatz}
{
	M^{[q]}(r,L_b,L)\propto L_b^{x_2(q)}L^{-y_2(q)}r^{-z(q)}
}
and try to find relations of the exponents $x_2(q)$, $y_2(q)$, $z(q)$
to the previous ones $x(q)$ and $y(q)$\footnote{In the following we
suppress the superscript $[Q]$ in exponents.}.
A typical heuristic scaling argument to fix scaling relations relies
on considering limiting cases where scaling will be already violated.
However, since the limit is reached continuously and scaling exponents are
unique one can conclude scaling relations due to consistency. We thus
proceed
by considering the limiting cases for $M^{[q]}$: (i) $r=L_b$ and (ii)
$r=L$.
In case (i) we expect the asymptotics
\equ{asy1}
{
	M^{[q]}\propto \BRA Q^{2q}(L_b)\KET_L\propto
L_b^{x(2q)}L^{-y(2q)}\, .
}
In case (ii) we expect the asymptotics
\equ{asy2}
{
	M^{[q]}\propto \BRA
\overline{Q^q}(L_b)\overline{Q^q}(L_b)\KET_L\propto \BRA
Q^q(L_b)\KET^2_L\propto L_b^{2x(q)}L^{-2y(q)}
}
where $\overline{Q^q}(L_b)$ is the value for $Q^q(L_b)$ already
averaged over one system of size $L$. Comparison with Eq.~(\ref{mansatz})
yields the scaling relations
\equa{scallla}
{
	y_2(q) &=& y(2q)\nonumber\\
	x_2(q) &=& x(2q) \nonumber\\
	z(q)&=& 2y(q)-y(2q)=2x(q)-x(2q)
}
which coincide with those of a more sophisticated
derivation by Cates and Deutsch$^{58}$ in the case
of $Q$ being a local measure. Notice  that the normalization exponents
$X$ and $Y$ cancel in $z(q)$ leading to
$z(q)=D+2\tau(q)-\tau(2q)$.
The analytic properties of $z(q)$,$x_2(q)$ and $y_2(q)$
are summarized as follows.
$z(q)$ vanishes at $q=0$ where it has a minimum. It is monotonically
increasing (decreasing) for $q>0$ ($q<0$), and it's slope vanishes for
$q\to\pm\infty$.
$x_2(q)$ and $y_2(q)$ have negative curvature. Their monotonicity
depend
on the sign of $\alpha(2q)+X$ and of $\alpha(2q)+Y$, respectively.
Since $\alpha(q)$ is monotonically decreasing they are monotonically
increasing
provided $D_{\infty}>-X,-Y$.
Notice that $x_2(q)$ and $y_2(q)$ depend on the normalization
exponents and may become negative for positive values of
$q$.

The  fact that $M^{[q]}$ can depend on the system size $L$
is unusual compared to ordinary critical phenomena where the
correlation function, the susceptibility $\chi(r)$, is expected to
depend on the distance $r$ only, provided the thermodynamic limit is
reached and the correlation length $\xi$ is infinite at the critical
point $T_c$. We like to comment on this in the following.

In Sec.~$3.2$ we discussed an interpretation of multifractality
of the box-probability in the context of the LD transition:
Multifractality reflects broadness of the box-probabilities
distribution function on all length scales which is due to the
dependence of the box-probability in each box on a large number of
conditions, simultaneously.
More general, we call a situation, {\em where a local box observable
depends on a large number of conditions for the entire system of
linear size $L$, simultaneously, a situation of  ``many parameter
(MP) coherence'' }.
In the context of the LD transition coherence at zero temperatures
is due to  quantum mechanical phase
coherence of the electrons wave function and disorder introduces a
huge number of parameters, e.g. the position of point-scatterers.
It may happen that MP coherence is valid only up to a certain scale
$L<{\hat{L}}$, defined implicitly by  $M^{[q]}$ being independent of
 $L$ for $L>{\hat{L}}$. We call $\hat{L}$ a MP coherence
length.
If such crossover in $M^{[q]}$ exists,  two situations have
to be distinguished. First, $\hat{L}$ introduces a cut-off for
correlations. For example,
the correlation length $\xi$ is a MP coherence length of this kind.
Alternatively, $\hat{L}$ does not introduce a
cut-off and correlations still show homogeneity exponents
for distances $r>{\hat{L}}$. An example for this kind of MP coherence
length in the LD transition problem will be discussed below.
In the latter situation the following scaling behavior of $M^{[q]}$
is expected to occur for $r\ll {\hat{L}}\ll L$,
\equa{conj1}
{
	M^{[q]}(r,L,{\hat{L}})\propto {\hat{L}}^{-y_2(q)}r^{-z(q)}\, ,
}
and for $L\gg r\gg {\hat{L}}$,
\equ{con2}
{
	M^{[q]}(r,L,{\hat{L}})\propto
	r^{-\tilde{z}(q)}\, ,
}
respectively.
Thus, for $r\gg {\hat L}$ the usual  behavior is recovered.
However, for this situation the multifractality on scales less than
${\hat{L}}$
is still reflected by the $q$-dependence of $\tilde{z}(q)$ the scaling
relation
of which can be concluded by a similar reasoning as that leading to
Eqs.~(\ref{scallla})
\equ{scalala}
{
	\tilde{z}(q)=y_2(q)+z(q)=2y(q)\, .
}
The analytic properties of $\tilde{z}(q)$ are such: $\tilde{z}(q)$ has
negative
curvature and its monotonicity properties depend on the sign of
$\alpha(q)+Y$.
It can thus happen, that $\tilde{z}(q)$ is negative for a wide range
of $q$ values.
A system which is never MP coherent, i.e. $\hat{L}$ is microscopic,
the single-fractal
situation
applies and, then,  $z(q)\equiv 0$ and $\tilde{z}(q)=2y(q)=y(2q)$.

\subsection{Application to LD Transitions}\label{subapp}
\noindent
In the LD transition problem the density of states of an individual
system is a suitable candidate for a local box-observable.
It is defined by
\equ{locde}
{
	\rho(E,\vec{r}):=\BRA\vec{r}|\delta (E-H)|\vec{r}\KET
=\sum_\alpha |\psi_\alpha(\vec{r})|^2 \delta(E-\varepsilon_\alpha)
}
where $\psi_\alpha$ are eigenstates of $H$ with respect to energy
$\varepsilon_\alpha$. One peculiar feature of the LD transition is
that the average density of states, $\rho(E)=\frac{1}{L^d}\Tr\delta(E-H)$,
is not an order parameter and is $L$-independent.
The corresponding box observable is
\equ{denbox}
{
	Q^{\rho}_i(L_b,E):=\int\limits_{\rm box (i)} d^d r \, \rho(E,\vec{r})
}
The scale-independence of $\rho(E)$
determines the normalization exponents to be
\equ{nrho}
{
	X^{\rho}=0\; ,\;\; Y^{\rho}=-D\, .
}
The fact that the average of $\rho(E)$ is not an order parameter ($\beta=0$)
is equivalent to $x^{\rho}(1)=0$. However, the typical value
of $Q_{\rm typ}^{\rho}$ can serve as an order parameter, because
\equ{orderja}
{
	Q_{\rm typ}^{\rho}\propto L^{-(\alpha_0-D)}
}
and consequently, by  $L$ approaching $\xi_c$, we have
\equ{ordja3}
{
	Q_{\rm typ}^{\rho}\propto t^{\nu\cdot(\alpha_0-D)}\, .
}
Thus, in
contrast to the mean value $\BRA Q^{\rho}\KET$,
$Q_{\rm typ}^{\rho}$ is able to indicate the LD transition and we have
reason to call
\equ{qtyp}
{
	 \beta_{\rm typ}:=\nu(\alpha_0+Y)=\nu(\alpha_0-D)
}
  the {\em
typical order parameter
 exponent}.

The exponents $z(q)$ do not depend on the normalization exponents
and are given as $z^{\rho}(q)=D+2\tau(q)-\tau(2q)$ whereas the exponents
$\tilde{z}^{\rho}(q)$
do depend on $Y^{\rho}=-D$,
\equ{tildez}
{
	\tilde{z}^{\rho}(q)=2(q-1)\LK D(q)-D\RK\,
}
have negative curvature, are positive for $0<q<1$, vanish at $q=0,1$ and
are negative elsewhere. The function $\tau(q)$ is, due to Eq.~(\ref{locde}),
the same as for the
wave function itself.

For $q=1$ we can  compare with the result of Chalker and Daniell$^{30}$
for the scaling of the
density correlator at the LD transition of a quantum Hall system,
\equ{chal1}
{
	\BRA \rho(E+\omega/2,\vec{r})\rho(E-\omega/2,\vec{r}\ ')\KET\, ,
}
where $\omega$ sets the scale of a MP coherence length
$L_\omega=(\omega\rho)^{-1/2}$. $L_\omega$ can be interpreted as the
linear system size for a system with mean level spacing of about
$\omega$.$^{30}$
Their result is: for $r\ll L_\omega$ correlations scale like
$r^{-0.38}$
whereas for $r\gg L_\omega$ correlations behave like
$r^{0}$. The first of these results was already discussed in
Secs.~$2$,$4$ and can be recovered
from Eq.~(\ref{scallla})
\equ{chalk3}
{
	z =z^{\rho}(1)=D-D(2)=0.38\, .
}
The second result is consistent with the interpretation of $\tilde{z}$
being the distance exponent in a regime $r\gg L_\omega$,
i.e.
\equ{challla}
{
	\tilde{z}=\tilde{z}^{\rho}(1)=0\, .
}
A similar observation was made in Ref.~$12$ for the LD transition
in  2-d system with spin-orbit coupling , though the
results were not conclusive concerning quantitative results for
$z^{\rho}(1)$.
Furthermore, the scaling of the density correlator
with respect to $L_\omega$ fulfills the scaling relation of
Eq.~(\ref{scallla}) which means here
\equ{challlll}
{
	y^{\rho}_2(1)=y^{\rho}(2)=-z=-0.38\, .
}
This came out in the work of Chalker et al. by studying the
combined variable $L_\omega k$ where $k$ is the inverse wavelength
in the Fourier-transform of the density correlator and probes long
 distances in the limit $k\to 0$.
Instead of interpreting $\omega$ via $L_\omega$ one can (more
directly)
look at
the density correlator of Eq.~(\ref{chal1}) as describing  the
density correlations in energy and, owing to Fourier transformation for
the limit $\omega\to 0$, the
long time correlations of wave packets. Huckestein and Schweitzer$^{50}$
 checked the latter point of view by explicitly calculating
the long time correlations of wave packets in the critical regime of a
quantum Hall system and found excellent agreement, i.e.
$z=0.38\pm0.02$.

Unfortunately, no results for values  $q\not=1$ for $z^{\rho}(q),
y^{\rho}_2(q)$
and $\tilde{z}^{\rho}(q) $ are available up to now. Thus,  the picture
developed here  needs further tests.

\subsection{Conformal Mapping in 2-d}\label{subcon}
\noindent
We  apply conformal mapping arguments  to find a
 relation between finite size scaling (FSS)
methods relying on strip-like systems and the multifractal analysis for
square-like systems in 2-d.
To begin with some ideas about conformal mapping are reviewed (for a
satisfactory treatment see e.g. Ref.~${59}$).

For critical correlation functions $A(r)\propto r^{-\kappa}$
the following   assumption seems
plausible. Scale invariance, as reflected by Eq.~(\ref{11.1}), should hold
also for local transformations which preserves angles but may change
scales locally in the sense that correlations of one system with a certain
geometry are mapped
onto those correlations of a similar system the geometry of which is
determined
 by the transformation. Such transformations are called conformal mappings.
The corresponding Jacobian $J$ has to fulfill
\equ{jak}
{
	\frac{\vec{v}\cdot \vec{w}}{\LK\vec{v}^2\vec{w}^2\RK^{1/2}} =
\frac{J\vec{v}\cdot
J\vec{w}}{\LK(J\vec{v})^2(J\vec{w})^2\RK^{1/2}}
}
where $\vec{v},\vec{w}$ are vectors of the tangent space at a given point.
In 2-d any holomorphic function $f(z=x+iy)$ is a conformal mapping.
Especially, the complex logarithm provides a conformal mapping of the
entire plane onto an infinite strip with periodic boundary conditions.
Let's introduce Cartesian $(x,y)$
and polar coordinates $(r,\varphi)$ on the plane by
$x+iy=re^{i\varphi}$.
Then the conformal mapping $F(x+iy)=\frac{M}{2\pi}\ln (x+iy)=:u+iv$
maps a ring area, denoted by ${\cal C}$,
with inner radius $1/L$ and outer radius $L$ onto a
strip
of width $M$ with length $L'=\frac{M}{\pi}\ln L$, centered at $u=0$ and
fulfilling periodic boundary conditions in $v$-direction.
For a  correlation function on the plane $(x,y)$,
$A(r)\propto r^{-\kappa}$ which behaves regular for $r\to 0$
we can express the exponent $\kappa$ by an integral expression in
coordinate free terms by (cf. Ref.~$60$)
\equ{kappa}
{
	\kappa=\frac{-1}{2\pi}\int\limits_{\cal C} \omega
}
where the two-form $\omega:=d*d(\ln A)$, $*$ is the Hodge-operator
corresponding to the Euclidean metric,
and $1/L$ is assumed to be small enough to give vanishing contribution
to $\kappa$ by applying Stokes' theorem.
By lifting $\omega$ with the conformal mapping $F$, and assuming
exponential decay of the corresponding correlation function along the strip,
$\tilde{A}(u,v)\propto \exp{-|u|/\xi(M)}$, one arrives at the result
\equ{result}
{
	\frac{\xi(M)}{M}=\frac{1}{\pi \kappa}\, .
}

We wish to apply this result to correlation functions of multifractal
correlations $M^{[q]}$ with $L_b$ being microscopic. However,
 in the regime $r\ll {\hat{L}},L$
conformal mapping arguments of the kind presented here can never apply
since a second length scale besides the distance $r$ appears.
We thus have to focus on the regime $r\gg {\hat{L}}$ where $M^{[q]}$
behaves as
\equ{mqphi}
{
	M^{[q]}(r)\propto r^{-\tilde{z}(q)}\, .
}
At the critical point of the LD transition, where power law in the
plane is valid, the decay lengths $\xi^{[q]}(M)$ of the corresponding
correlations in the strip-geometry vary linear with $M$. This fact is
exploited
in the FSS analysis of the LD transition (see e.g. Ref.~$14$).
Thus, the FSS variable $\Lambda^{[q]}(M):=
2\xi^{[q]}(M)/M$\footnote{The factor of 2 is due to convention.}
becomes a constant $\Lambda^{[q]}_c$ at the critical point.

Making the hypothesis, that the conformal mapping result
Eq.~(\ref{result})
applies to the multifractal correlations in the regime described by
Eq.~(\ref{mqphi}) one would conclude
\equ{con1}
{
	\Lambda^{[q]}_c=2/(\pi \tilde{z}(q))\, .
}
However, in the situation of the LD transition
$\tilde{z}(q)$
takes negative values for $q>1$ which is counterintuitive and suggests
exponential growth of correlations in the strip geometry. Furthermore,
the exponents $\tilde{z}(q)$ and the decay lengths $\xi^{[q]}$ are
attached to averages of powers of the correlation function, though
the precise statement on the existence of the decay length in the
strip-geometry relies on the average of the logarithms of the
correlations (see e.g. Ref.~$14$).
Thus, we leave the validity of Eq.~(\ref{con1}) as an open question
and turn over to the typical values defined by the average of logarithms.
In FSS calculations it is the average of the logarithm of the
correlation function which is calculated by knowing that this quantity
is self-averaging. Therefore, we denote results of such FSS
calculations
 by $\Lambda^{\rm typ}_c$ and
a conjecture about the relation to multifractality (relying on
conformal mapping arguments) reads
\equ{dbfc}
{
	\Lambda^{\rm typ}_c=2/(\pi \tilde{z}_{\rm typ})\; ,\;\;
\tilde{z}_{\rm typ}=2(\alpha_0 + Y)}
where the expression for $\tilde{z}_{\rm typ}$ is
 a consequence of Eqs.~(\ref{scalala},\ref{addob}).

In the context of the LD transition we found $Y^{\rho}=-D$ for the local
observable being the density of states.  This observable involves
the square amplitude of  wave functions and localization lengths are
usually
defined with respect to the modulus of the wave function. This explains the
conventional factor of $2$ in the definition of $\Lambda(M)$.

Up to now, only data for two universality classes are known where one
can  check
 the
prediction
of Eq.~(\ref{dbfc}): (i) quantum Hall systems and (ii)
2-d Anderson model with spin-orbit scattering.
Taking into account that present days calculations, lacking  larger
systems,
are able to produce results for $\Lambda^{\rm typ}_c$ and $\alpha_0-D$
with a precision  hardly exceeding $5\%$, the results are not in
conflict
with Eq.~(\ref{dbfc}): (i) $\Lambda^{\rm typ}_c=1.14$$^{61}$ and
$1/(\pi(\alpha_0-2))=1.06$.$^{43}$ (ii) $\Lambda^{\rm
typ}=1.88$$^{62}$ and $1/(\pi(\alpha_0-2))=1.82$.$^{57}$

\section{Conclusions}
\noindent
After reminding some aspects of critical phenomena in Sec.~$2$
we described the multifractal analysis of broadly distributed
observables in the critical regime of a critical phenomenon. Broadness
of the distribution on length scales much less than the correlation
length $\xi_c$  is determined by the single-humped $f(\alpha)$
spectrum
of normalized box-observables. The parabolic approximation
for $f(\alpha)$ corresponds to a log-normal distribution.
 Scaling of the $q$-th moment is given by
the function $\tau(q)$ defining generalized dimensions
$D(q)=\tau(q)/(q-1)$ ($D=D(0)$ being the geometric fractal dimension)
and by the normalization exponents $X,Y$.
The function $\tau(q)$ is monotonically  increasing and has negative
curvature.
It is related to $f(\alpha)$ by Legendre transformation (Secs.~$3-4$).
Since the distribution is broad it is not possible to characterize
it by the mean value. Instead, the typical value defined as the
geometric mean is a self-averaging quantity which scales with exponent
$\alpha_0+X$ where $\alpha_0$ is the maximum position of $f(\alpha)$,
$f(\alpha_0)=D$. We interpreted the appearence of broad distributions
being due to {\em many parameter coherence}, saying that local
observables
depend on a large number of conditions for the entire system, simultaneously.

Furthermore, correlation functions of broadly distributed observables
at criticallity show scaling dependence with respect to the distance
between
local observables {\em and} with respect to the linear system size
$L$
or, equivalently, with respect to a length $\hat{L}$ indicating the
range
of many parameter coherence. For distances $r$ exceeding the length
$\hat{L}$ correlations will only depend on $r$ (Sec.~$5$). Although
compatible with known results,  this
scenario needs further investigations.

Scaling exponents of correlations are related to
$\tau(q)$, $X$, $Y$ by scaling relations (Eqs.~(\ref{scallla},\ref{scalala}))
Therefore, a complete characterization of critical exponents in terms
of multifractal spectra is possible.

We applied the multifractal analysis to the
localization-delocalization   transition induced by disorder
(LD transition) leading to the following
conclusions. The formal order parameter, the average density of
states,
does not show the LD transition which means the order parameter
exponent vanishes,
$\beta=0$. Consequently,
usual scaling relations tell that correlations show a distance
exponent
$\tilde{z}=0$. In contrast to the average density of states, the
  typical value is able to reflect the LD transition and the
corresponding
typical order parameter exponent and distance exponent are $\beta_{\rm
typ}=(\alpha_0-D)\nu$ and $\tilde{z}_{\rm typ}=2(\alpha_0-D)$,
respectively (Sec.~$5$). Here $\nu$ is the critical exponent of the
localization length. By choosing the typical conductance as a scaling
variable
an expression for $\nu$ in terms of only multifractal exponents can be
given, $\nu=(\alpha_0+X^{[g]})^{-1}$. The multifractal analysis
yields lower as well as upper bounds for $\nu$ in terms of $D(2)$ and
$\alpha_0$, respectively (Eqs.~(\ref{11.73}, \ref{11.78})).
The  exponent $z$ describing the distance exponent in the regime $r\ll
L, {\hat{L}}$  is given by $z=d-D(2)\not=\tilde{z}$
 which is excellently confirmed
for
the LD transition in quantum Hall systems (Sec.~$5$).

With the help of conformal mapping arguments in $d=2$ we suggested a
relation between the critical value of the (typical) renormalized
localization
length $\Lambda_c$ of strip-like systems  (being the scaling variable
 in finite size scaling calculations) and the exponent $\alpha_0$
(Eq.~(\ref{dbfc})).
 This relation is in agreement with presently available data.
A corresponding relation between moments (Eq.~(\ref{con1}))
 needs further numerical
investigations.
Such relation is highly desirable since it would allow to demonstrate
complete
equivalence
between the multifractal analysis and the finite size scaling
approach.

Finally, we comment on the question of how to determine the critical
point $T_c$ by the multifractal analysis. For finite system sizes
$L$ the critical regime of values $T$ around $T_c$, characterized by
$L\ll \xi_c$, is finite. Within this regime the $f(\alpha)$ spectra
can be calculated by determining linear regimes in $\log$-$\log$ plots
(see Eqs.~(\ref{11.26})). These values will slightly differ from
the universal values at $T_c$. The difference (finite size
corrections)
will scale, e.g.$^{63}$
$\alpha_0(T)-\alpha_0(T_c) \propto |T-T_c|L^{1/\nu}$.
Thus, in principle, one can determine $T_c$ and $1/\nu$ from finite
size corrections.$^{17}$ There have been speculations$^{64,65}$ to determine
$T_c$, in the context of the LD transition, by requiring a certain  fractal
dimension to coincide with the lower critical dimension $d_l$ which
is believed to be $d_l=2$. Since it is known that there is a spectrum
of fractal dimensions, such criteria are not evident.
To the authors knowledge the only motivation for  conjecturing
$D(1)=d_l$ as a criterium to fix $T_c$  comes
from the one-loop result of Wegners non-linear $\sigma$-model in
$d=2+\epsilon$$^{19,47}$ where
$D(q)=2+\epsilon -q\epsilon$. However, this approximation is expected
to deviate from exact values as $\epsilon \to 1$.
Thus, in order to determine $T_c$ by the multifractal analysis
one has to establish universality in multifractal exponents rather
than to establish certain values for these exponents.

\nonumsection{Acknowledgements}
\noindent
This work was supported by the Deutsche Forschungsgemeinschaft.
I thank the Department of Theoretical Physics in Oxford, U.K., for the
kind hospitality during a visit in which most part of this work was
performed.
I gratefully acknowledge helpful discussions with J. Chalker, U.
Fastenrath,
J. Hajdu, B.
Huckestein, W. Pook, L. Schweitzer and M. Zirnbauer.

\nonumsection{References}

\end{document}